\documentclass[journal]{IEEEtran}\newcommand{\figsize}{3.5}
\IEEEoverridecommandlockouts
\usepackage{graphicx,psfrag}
\usepackage{url,hyperref}
\usepackage[usenames]{color}
\usepackage[cmex10]{amsmath}
\usepackage{amsfonts,amssymb,latexsym} 

{\global\def\bigPage{
        \setlength{\topmargin}{-0.5in}
        \setlength{\textheight}{9.25in}
        \setlength{\oddsidemargin}{-0.25in}
        \setlength{\textwidth}{7.0in}
        }
}

 \newcommand{\putFrag}[4]{\begin{figure}[h]
                            \centering
                            #4
			    \includegraphics[width=#3in]{figures/#1.eps}
            		    \caption{#2}
           		    \label{fig:#1}
                          \end{figure} }
 \newcommand{\putTable}[3]{\begin{table}[t!]
  			    \centering
		            #3
     			    \caption{#2}
     			    \label{tab:#1}
			  \end{table} }

 \renewcommand{\tilde}{\widetilde}
 \newcommand{\defn}{\triangleq}

 \newcommand{\ovec}[1]{\ensuremath{\Bar{\boldsymbol{#1}}}}
 \newcommand{\hvec}[1]{\ensuremath{\Hat{\boldsymbol{#1}}}}
    
 \renewcommand{\vec}[1]{\ensuremath{\boldsymbol{#1}}}
 \newcommand{\mat}[1]{\ensuremath{\begin{bmatrix}#1\end{bmatrix}}}

 \newcommand{\norm}[1]{\ensuremath{\| #1 \|}}
 \newcommand{\mc}[1]{\ensuremath{\mathcal{#1}}}

 \newcommand{\Real}{{\mathbb{R}}}
 \newcommand{\Complex}{{\mathbb{C}}}

 \newcommand{\tran}{^\textsf{T}}
 \newcommand{\herm}{^\textsf{H}}

 \renewcommand{\hat}[1]{\widehat{#1}}

 \DeclareMathOperator{\real}{Re}
 \DeclareMathOperator{\imag}{Im}

 \DeclareMathOperator{\E}{E}
 \DeclareMathOperator{\var}{var}

 \renewcommand{\eqref}[1]{(\ref{eq:#1})}
 
 \newcommand{\Figref}[1]{Figure~\ref{fig:#1}}
 \newcommand{\figref}[1]{Fig.~\ref{fig:#1}}
 \newcommand{\tabref}[1]{Table~\ref{tab:#1}}
 \newcommand{\secref}[1]{Section~\ref{sec:#1}}
 
 \newcommand{\appref}[1]{Appendix~\ref{app:#1}}

 \newcommand{\textb}[1]{#1}

 \newcounter{comment}[section]
 
 \newcounter{texthead}[section]

\newcommand{\SNR}{\textsf{SNR}}
\newcommand{\NMSE}{\textsf{NMSE}}

\newcommand{\inp}{_{\textsf{in},n}}
\newcommand{\out}{_{\textsf{out},m}}

\newcommand{\NRES}{\textsf{NR}}
\newcommand{\best}{_{\textsf{best}}}
\newcommand{\init}{_{\textsf{init}}}
\newcommand{\suff}{_{\textsf{stop}}}
\newcommand{\true}{_{\textsf{true}}}
\newcommand{\dB}{{\textsf{dB}}}

\begin{document}
\bigPage
\setlength{\arraycolsep}{0.8mm}
 \title{Compressive Phase Retrieval via Generalized Approximate Message Passing}
         \author{Philip Schniter\IEEEauthorrefmark{1} and Sundeep Rangan
         \thanks{Philip Schniter is with the Department of Electrical and Computer Engineering, The Ohio State University, Columbus OH and 
		 Sundeep Rangan is with the Department of Electrical and Computer Engineering, New York University, Brooklyn, NY.}
	 \thanks{\IEEEauthorrefmark{1}Please direct all correspondence to Prof. Philip Schniter, Dept. ECE, The Ohio State University, 2015 Neil Ave., Columbus OH 43210, e-mail: schniter@ece.osu.edu, phone 614.247.6488, fax 614.292.7596.}
         \thanks{This work was supported in part by NSF grant CCF-1018368, by DARPA/ONR grant N66001-10-1-4090, and by an allocation of computing time from the Ohio Supercomputer Center.}
	 \thanks{Portions of this work were presented at the Allerton Conference in Oct.~2012 \cite{Schniter:ALL:12} and the FFT Workshop in Feb.~2013 \cite{Schniter:FFT:13}.}
        }%
 \date{\today}
 \maketitle

\begin{abstract}
In phase retrieval, the goal is to recover a signal $\vec{x}\in\mathbb{C}^N$ from the magnitudes of linear measurements $\vec{Ax}\in\mathbb{C}^M$.
While recent theory has established that $M\approx 4N$ intensity measurements are necessary and sufficient to recover generic $\vec{x}$, there is great interest in reducing the number of measurements through the exploitation of sparse $\vec{x}$, which is known as compressive phase retrieval.
In this work, we detail a novel, probabilistic approach to compressive phase retrieval based on the generalized approximate message passing (GAMP) algorithm.
We then present a numerical study of the proposed PR-GAMP algorithm, demonstrating its excellent phase-transition behavior, robustness to noise, and runtime.
\textb{Our experiments suggest that approximately $M\geq 2K\log_2(N/K)$ intensity measurements suffice to recover $K$-sparse \textb{Bernoulli-Gaussian} signals for $\vec{A}$ with i.i.d Gaussian entries and $K\ll N$. Meanwhile,}
when recovering a 6k-sparse 65k-pixel grayscale image from 32k randomly masked and blurred Fourier intensity measurements \textb{at 30~dB measurement SNR}, PR-GAMP achieved \textb{an output SNR of no less than 28~dB in all of 100 random trials,} with a median runtime of only \textb{7.3} seconds.
Compared to the recently proposed CPRL, sparse-Fienup, and GESPAR algorithms, our experiments \textb{suggest} that PR-GAMP has a superior phase transition and orders-of-magnitude faster runtimes as the \textb{sparsity and} problem dimensions increase.
\end{abstract}


\section{Introduction} \label{sec:intro}

\subsection{Phase retrieval}

In phase retrieval, the goal is to recover a signal $\vec{x}\in\Complex^N$ from the \emph{magnitudes} $y_m=|u_m|$ of possibly noisy linear measurements $\vec{u}=[u_1,\dots,u_M]\tran=\vec{Ax}+\vec{w}\in\Complex^M$. 
This problem is motivated by the fact that it is often easier to build detectors (e.g., photographic plates or CCDs) that measure intensity rather than phase \cite{Fienup:AO:82,Millane:SPIE:06}.
Imaging applications of phase retrieval include
X-ray diffraction imaging \cite{Bunk:AC:07},
X-ray crystallography \cite{Harrison:JOSAA:93,Millane:JOSAA:90},
array imaging \cite{Chai:IP:11},
optics \cite{Walther:OA:63},
speckle imaging in astronomy \cite{Dainty:Chap:87},
and microscopy \cite{Miao:ARPC:08}.
Non-imaging applications include
acoustics \cite{Balan:ACHA:06},
interferometry \cite{Demanet:13}, and
quantum mechanics \cite{Corbett:RMP:06}.

To reconstruct $\vec{x}\in\Complex^N$ (up to a global phase uncertainty), it has been recently established that 
$M\geq 4N-o(N)$ intensity measurements are necessary \cite{Heinosaari:CMP:13} and $M\geq 4N-4$ are sufficient \cite{Bodmann:ACM:14} through appropriate design of the linear transform $\vec{A}$.
Meanwhile, to reconstruct $\vec{x}\in\Real^N$ (up to a global sign uncertainty), it has been shown that $M\geq 2N-1$ measurements are both necessary and sufficient \cite{Balan:ACHA:06}.
However, there exist applications where far fewer measurements are available, such as
sub-wavelength imaging \cite{Gazit:OX:09,Szameit:NM:12},
Bragg sampling from periodic crystalline structures \cite{Marchesini:08},
and waveguide-based photonic devices \cite{Shechtman:OX:13}.
To facilitate these \emph{compressive} phase retrieval tasks, it has been proposed to exploit \emph{sparsity}\footnote{%
$\vec{x}$ may represent the sparse transform coefficients of a non-sparse signal-of-interest $\vec{s}=\vec{\Psi x}$ in a sparsifying basis (or frame) $\vec{\Psi}$, in which case the intensity measurements would be $\vec{y}=|\vec{\Phi s}+\vec{w}|$ and $\vec{A}\defn\vec{\Phi\Psi}$.
} in $\vec{x}$.
In fact, very recent theory confirms the potential of this approach:
to reconstruct $K$-sparse $N$-length $\vec{x}$ using a generic (e.g., i.i.d Gaussian) $\vec{A}$, only $M\geq4K-2$ intensity measurements suffice in the complex case and $M\geq2K$ suffice in the real case (where $M\geq2K$ is also necessary) when $K<N$ \cite{Li:JMA:13}. 
While these bounds are extremely encouraging, achieving them with a practical algorithm remains elusive.

To our knowledge, the first algorithm for compressive phase retrieval was proposed by Moravec, Romberg, and Baraniuk in \cite{Moravec:SPIE:07} and worked by incorporating an $\ell_1$-norm \emph{constraint} into a traditional Fienup-style \cite{Fienup:AO:82} iterative algorithm.
However, this approach requires that the $\ell_1$ norm of the true signal is known, which is rarely the case in practice.
Recently, a more practical sparse-Fienup algorithm was proposed by Mukherjee and Seelamantula \cite{Mukherjee:TSP:14}, which requires knowledge of only the signal sparsity $K$ but is applicable only to measurement matrices $\vec{A}$ for which $\vec{A}\herm\vec{A}=\vec{I}$.
Although this algorithm guarantees that the residual error $\|\vec{y}-|\vec{A}\hvec{x}(t)|\|_2^2$ is non-increasing over the iterations $t$, it succumbs to local minima and, as we show in \secref{gespar}, is competitive only in the highly sparse regime.

To circumvent the local minima problem, Ohlsson, Yang, Dong, and Sastry proposed the \emph{convex relaxation} known as Compressive Phase Retrieval via Lifting (CPRL) \cite{Ohlsson:NIPS:12}, which adds $\ell_1$ regularization to the well-known PhaseLift algorithm \cite{Chai:IP:11,Candes:CPAM:13}.
Both CPRL and PhaseLift ``lift'' the unknown vector $\vec{x}\in\Complex^N$ into the space of $N\times N$ rank-one matrices and solve a semidefinite program in the lifted space, requiring $O(N^3)$ complexity, which is impractical for practical image sizes $N$.
Subsequent theoretical analysis \cite{Li:JMA:13} revealed that, while $M\gtrsim O(K^2 \log N)$ intensity measurements suffice for CPRL when $\vec{x}\in\Real^N$, $M\gtrsim O(K^2/\log^2 N)$ measurements are \emph{necessary}, which is disappointing because this greatly exceeds the $2K$ measurements that suffice for the optimal solver \cite{Li:JMA:13}.
\textb{That said, the noise-robustness of PhaseLift-type algorithms and the sufficiency of $M\gtrsim O(K\log^2 N)$ samples for $K$-sparse signals with power-law decay has been recently established \cite[Thm.~4]{Chen:13}.
Also}, a cleverly initialized alternating minimization (AltMin) approach was recently proposed by Natrapalli, Jain, and Sanghavi in \cite{Netrapalli:NIPS:13} that gives CPRL-like guarantees/performance with only $O(NK^3)$ complexity.  However, this is still too complex for practical sparsities $K$, which tend to grow linearly with image size $N$.

Recently, Shechtman, Beck, and Eldar proposed the GrEedy Sparse PhAse Retrieval (GESPAR) algorithm \cite{Shechtman:TSP:14}, which applies fast 2-opt local search \cite{Papadimitriou:Book:98} to a sparsity constrained non-linear optimization formulation of the phase-retrieval problem.
Numerical experiments (see \secref{gespar}) \textb{suggest} that GESPAR handles higher sparsities $K$ than the sparse-Fienup technique from \cite{Mukherjee:TSP:14}, but at the cost of significantly increased runtime. 
In fact, due to the combinatorial nature of GESPAR's support optimization, its complexity scales very rapidly in $K$, making it impractical for many problems of interest.

In this work, we describe a novel\footnote{We described an earlier version of PR-GAMP in the conference paper \cite{Schniter:ALL:12} and the workshop presentation \cite{Schniter:FFT:13}.} approach to compressive retrieval that is based on loopy belief propagation and, in particular, the \emph{generalized approximate message passing} (GAMP) algorithm from \cite{Rangan:ISIT:11}.
In addition to describing and deriving our phase-retrieval GAMP (PR-GAMP) algorithm, we present a detailed numerical study of its performance.
For i.i.d Gaussian, Fourier, and masked-Fourier matrices $\vec{A}$, we demonstrate that PR-GAMP performs far better than existing compressive phase-retrieval algorithms in terms of both success rate and runtime for large values $K$ and $N$.
\textb{Our experiments suggest that} PR-GAMP requires approximately $4\times$ the number of measurements as phase-oracle GAMP (i.e., GAMP given the magnitude-and-phase measurements $\vec{u}=\vec{Ax}+\vec{w}$). 
\textb{Interestingly, for \emph{non}-sparse signals in $\Complex^N$, the}
ratio of magnitude-only to magnitude-and-phase measurements necessary and sufficient for perfect recovery is also known to be $4\times$ \textb{(as $N\rightarrow\infty$)} \cite{Heinosaari:CMP:13,Bodmann:ACM:14}.
\textb{Our experiments also suggest that} PR-GAMP is robust to additive noise, giving mean-squared error that is only $3$~dB worse than phase-oracle GAMP over a wide SNR range.

\emph{Notation}:
  For matrices, we use boldface capital letters like $\vec{A}$, and we use $\vec{A}\tran$, $\vec{A}\herm$, and $\|\vec{A}\|_F$ to denote the transpose, Hermitian transpose, and Frobenius norm, respectively.
  For vectors, we use boldface small letters like $\vec{x}$, and we use $\|\vec{x}\|_p=(\sum_n |x_n|^p)^{1/p}$ to denote the $\ell_p$ norm, with $x_n=[\vec{x}]_n$ representing the $n^{th}$ element of $\vec{x}$.
  For random variable $X$, we write the pdf as $p_{X}(x)$, the expectation as $\E\{X\}$, and the variance as $\var\{X\}$.
  In some cases where it does not cause confusion, we drop the subscript on $p_{X}(x)$ and write the pdf simply as $p(x)$.
  For a ``circular Gaussian'' random variable $X\textb{\in\Complex}$ with mean $m$ and variance $v$, we write the pdf as $p_X(x)=\mc{N}(x;m,v)\defn\frac{1}{\pi v} \exp(-|x-m|^2/v)$.
  \textb{Note that $X\sim\mc{N}(m,v)$ has real and imaginary components that are jointly Gaussian, uncorrelated, and of equal variance $v/2$.}
  For the point mass at $x=0$, we use the Dirac delta distribution $\delta(x)$. 
  Finally,
  we use $\Real$ for the real field,
  $\Complex$ for the complex field, 
  $\textrm{Re}\{x\}$ and $\textrm{Im}\{x\}$ for the real and imaginary parts of $x$,
  and $x^*$ for the complex conjugate of $x$.


\section{Background on GAMP}

The approximate message passing (AMP) algorithm was recently proposed by Donoho, Maleki, and Montanari \cite{Donoho:PNAS:09,Donoho:ITW:10a} for the task estimating a signal vector $\vec{x}\in\Real^N$ from linearly transformed and additive-Gaussian-noise corrupted measurements\footnote{Here and elsewhere, we use $\vec{y}$ when referring to the $M$ measurements that are available for signal reconstruction.  
In the canonical (noisy) compressive sensing problem, the measurements take the form $\vec{y} = \vec{Ax}+\vec{w}$, but in the (noisy) compressive phase retrieval problem, the measurements instead take the form $\vec{y} = |\vec{Ax}+\vec{w}|$.}
\begin{equation}
\vec{y} = \vec{Ax}+\vec{w}\in\Real^M .	\label{eq:y} 
\end{equation}
The Generalized-AMP (GAMP) algorithm proposed by Rangan \cite{Rangan:ISIT:11} then extends the methodology of AMP to the generalized linear measurement model   
\begin{equation}
\vec{y} = q(\vec{Ax}+\vec{w})\in\Real^M ,	\label{eq:yq} 
\end{equation}
where $q(\cdot)$ is a component-wise nonlinearity.
This nonlinearity facilitates the application of AMP to phase retrieval.

Both AMP and GAMP can be derived from the perspective of \emph{belief propagation} \cite{Pearl:Book:88}, a Bayesian inference strategy that is based on a factorization of the signal posterior pdf $p(\vec{x}|\vec{y})$ into a product of simpler pdfs that, together, reveal the probabilistic structure in the problem.
Concretely, if we model the signal coefficients in $\vec{x}$ and noise samples in $\vec{w}$ from \eqref{y}-\eqref{yq} as statistically independent, so that $p(\vec{x})=\prod_{n=1}^N p_{X_n}\!(x_n)$ and $p(\vec{y}|\vec{z})=\prod_{m=1}^M p_{Y|Z}(y_m|z_m)$ for $\vec{z}\defn\vec{Ax}$, then we can factor the posterior pdf as
\begin{align}
 p(\vec{x}|\vec{y})
 &\propto p(\vec{y}|\vec{x}) p(\vec{x}) \\
 &= \prod_{m=1}^M p_{Y|Z}\big(y_m\big| [\vec{Ax}]_m\big) 
           \prod_{n=1}^N p_{X_n}\!(x_n) ,            
 \nonumber\\[-4mm]&\label{eq:decoupled}
\end{align}
yielding the factor graph in \figref{fg_iid}.

\putFrag{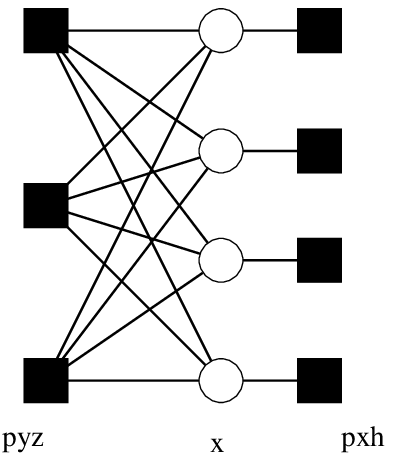}
   {GAMP factor graph, with white circles denoting random variables and black squares denoting pdf factors, for the case $M=3$ and $N=4$.}
   {1.2}
   {\newcommand{\sz}{0.9}
    \psfrag{pyz}[Br][br][\sz]{$p_{Y|Z}(y_m|\sum_{n}a_{mn}x_n)$}
    \psfrag{pxh}[Bl][bl][\sz]{$p_{X_n}\!(x_n)$}
    \psfrag{x}[B][b][\sz]{$x_n$}}

In belief propagation \cite{Pearl:Book:88}, beliefs about the unknown variables are passed among the nodes of the factor graph until all agree on a common set of beliefs.
The set of beliefs passed into a given variable node are then used to determine the posterior pdf of that variable, or an approximation thereof.
The sum-product algorithm \cite{Kschischang:TIT:01} is perhaps the most well-known incarnation of belief propagation,
wherein the messages take the form of pdfs and exact posteriors are guaranteed whenever the graph does not have loops.
For graphs with loops, exact inference is known to be NP hard, and so loopy belief propagation (LBP) is not guaranteed to produce correct posteriors.
Still, LBP has shown state-of-the-art performance on many problems in, e.g., decoding, computer vision, and compressive sensing \cite{Frey:NIPS:97}.

The conventional wisdom surrounding LBP says that accurate inference is possible only when the circumference of the loops are relatively large.
With \eqref{y}-\eqref{yq}, this would require that $\vec{A}$ is a sparse matrix, which precludes most interesting cases of compressive inference, including compressive phase retrieval.
Hence, the recent realization by Donoho, Maleki, Montanari, and Bayati that LBP-based compressive sensing is not only feasible \cite{Donoho:PNAS:09,Donoho:ITW:10a} for \emph{dense} matrices $\vec{A}$, but provably accurate \cite{Bayati:TIT:11,Bayati:ISIT:12}, was a breakthrough.
In particular, they established that, in the large-system limit (i.e., as $M,N\rightarrow\infty$ with $M/N$ fixed) and under i.i.d sub-Gaussian $\vec{A}$, the iterations of AMP are governed by a state-evolution whose fixed points describe the algorithm's performance.
To derive the AMP algorithm, \cite{Donoho:PNAS:09,Donoho:ITW:10a} proposed an ingenious set of message-passing approximations that become exact in the limit of large sub-Gaussian $\vec{A}$. 

Remarkably, the ``approximate message passing'' (AMP) principles in \cite{Donoho:PNAS:09,Donoho:ITW:10a}---including the state evolution---can be extended from the linear model \eqref{y} to the generalized linear model in \eqref{yq}, as established in \cite{Rangan:ISIT:11}.
The GAMP algorithm from \cite{Rangan:ISIT:11} is summarized in \tabref{gamp}. 
\textb{It is possible to recover the Bayesian AMP algorithm \cite{Donoho:ITW:10a} from \tabref{gamp} by considering the special case of $g\out(\hat{p},\nu^p)=(y_m-\hat{p})/(\nu^p+\nu^w)$ in line (D2) and by replacing all terms $|a_{mn}|^2$ in lines (R1) and (R5) with the constant value $M^{-1}$ (assuming that $\|\vec{A}\|_F^2=N$).
In the AMP literature, the term $\nu^p_m(t)\hat{s}_m(t\!-\!1)$ in (R2) is often referred to as the ``momentum'' or ``Onsager correction'' term.}

As in \cite{Rangan:ISIT:11}, we state the GAMP algorithm in a way that facilitates the use of complex-valued quantities, which is the case of interest in phase retrieval.
However, we note that the GAMP algorithm as stated in \tabref{gamp} is fully justified only in the case that all Gaussian random variables are circular (i.e., having independent real and imaginary components with identical variances),
and we use $\mc{N}(z;\hat{z},\nu^z)$ to denote the circular-Gaussian pdf in variable $z$ with mean $\hat{z}$ and variance $\nu^z$.
In the sequel, we detail how GAMP allows us to tackle the compressive phase retrieval problem.

\putTable{gamp}{The GAMP Algorithm from \cite{Rangan:ISIT:11} with $T_{\max}$ iterations.}{\footnotesize
\begin{equation*}
\begin{array}{|lrcl@{\;}r|}\hline
  \multicolumn{5}{|l|}{\texttt{input 
  	$\vec{A},
	 \{p_{X_n}\!(\cdot),\hat{x}_n(1),\nu^x_n(1)\}_{n=1}^N,
	 \{p_{Y|Z}(y_m|\cdot),\hat{s}_m(0)\}_{m=1}^M$
	 }}\\[1mm]
  \multicolumn{5}{|l|}{\texttt{define}}\\[-1mm]
  &p_{Z|Y,P}(z|y,\hat{p};\nu^p)
   &=& \frac{p_{Y|Z}(y|z) \,\mc{N}(z;\hat{p},\nu^p)}
        {\int_{\Complex} p_{Y|Z}(y|z') \,\mc{N}(z';\hat{p},\nu^p) dz'} &\text{(D1)}\\
  &g\out(\hat{p},\nu^p)
   &=& \frac{1}{\nu^p} \big(\E_{Z|Y,P}\{ Z|y_m,\hat{p};\nu^p\}-\hat{p}\big) &\text{(D2)}\\
  &g\out'(\hat{p},\nu^p)
   &=& \frac{1}{\nu^p}\Big(\frac{\var_{Z|Y,P}\{Z|y_m,\hat{p};\nu^p\}}{\nu^p}-1\Big)&\text{(D3)}\\
  &p_{X_n|R_n}(x|\hat{r};\nu^r)
   &=& \frac{p_{X_n}\!(x) \,\mc{N}(x;\hat{r},\nu^r)}
        {\int_{\Complex}p_{X_n}\!(x') \,\mc{N}(x';\hat{r},\nu^r)dx'}&\text{(D4)}\\
  &g\inp(\hat{r},\nu^r)
   &=& \E_{X_n|R_n}\{X_n|\hat{r};\nu^r\} &\text{(D5)}\\
  &g\inp'(\hat{r},\nu^r)
   &=& \var_{X_n|R_n}\{X_n|\hat{r};\nu^r\}\quad &\text{(D6)}\\[2mm]
  \multicolumn{5}{|l|}{\texttt{for $t=1,2,3,\dots,T_{\max}$}}\\
  &\forall m:
   \nu^p_m(t)
   &=& \textstyle \sum_{n=1}^{N} |a_{mn}|^2 \nu^x_n(t) & \text{(R1)}\\
  &\forall m:
   \hat{p}_m(t)
   &=& \textstyle \sum_{n=1}^{N} \!a_{mn} \hat{x}_n(t) - \nu^p_m(t) \,\hat{s}_m(t\!-\!1)& \text{(R2)}\\
  &\forall m:
   \hat{s}_m(t)
   &=& g\out(\hat{p}_m(t),\nu^p_m(t)) & \text{(R3)}\\
  &\forall m:
   \nu^s_m(t)
   &=& -g'\out(\hat{p}_m(t),\nu^p_m(t)) & \text{(R4)}\\
  &\forall n:
   \nu^r_n(t)
   &=& \textstyle \big(\sum_{m=1}^{M} |a_{mn}|^2 \nu^s_m(t) 
        \big)^{-1} & \text{(R5)}\\
  &\forall n:
   \hat{r}_n(t)
   &=& \textstyle \hat{x}_n(t)+ \nu^r_n(t) \sum_{m=1}^{M} \!a_{mn}^*
        \hat{s}_m(t)  & \text{(R6)}\\
  &\forall n:
   \nu^x_n(t\!+\!1)
   &=& \nu^r_n(t) g'\inp(\hat{r}_n(t),\nu^r_n(t)) & \text{(R7)}\\
  &\forall n:
   \hat{x}_n(t\!+\!1)
   &=& g\inp(\hat{r}_n(t),\nu^r_n(t)) & \text{(R8)}\\
  \multicolumn{5}{|l|}{\texttt{end}}\\[1mm]
  \multicolumn{5}{|l|}{\texttt{output 
  	$\{\hat{x}_n(T_{\max}\!+\!1),\nu^x_n(T_{\max}\!+\!1)\}_{n=1}^N,
	 \{\hat{s}_m(T_{\max})\}_{m=1}^M$
	 }}\\\hline
\end{array}
\end{equation*}
}

\section{Phase Retrieval GAMP}	\label{sec:PR-GAMP}

To apply the GAMP algorithm outlined in \tabref{gamp} to compressive phase retrieval, we specify a measurement likelihood function $p_{Y|Z}(y_m|\cdot)$ that models the lack of phase information in the observations $y_m$ and a signal prior pdf $p_{X_n}\!(\cdot)$ that facilitates measurement compression, e.g., a sparsity-inducing pdf.
In addition, we propose several extensions to the GAMP algorithm that aim to improve its robustness, and we propose an expectation-maximization method to learn the noise variance that parameterizes $p_{Y|Z}(y_m|\cdot)$.

\subsection{Likelihood function}	\label{sec:likelihood}
Before deriving the likelihood function $p_{Y|Z}(y_m|\cdot)$, we introduce some notation.
First, we will denote the noiseless transform outputs by 
\begin{align}
  z_m 
  &\defn \vec{a}_m\herm \vec{x} 
  = |z_m| e^{j\phi_m}
  \text{~with~} \phi_m\in[0,2\pi),	\label{eq:zm}
\end{align}
where $\vec{a}\herm_m$ is the $m$th row of $\vec{A}$
and $j\defn\sqrt{-1}$.  
Next, we will assume the presence of additive 
noise $w_m$ and denote the noisy transform outputs by 
\begin{align}
  u_m 
  &\defn z_m + w_m 
  = |u_m| e^{j\theta_m} 
  \text{~with~} \theta_m\in[0,2\pi).
\end{align}
Our (noisy) intensity measurements are then
\begin{align}
  y_m
  &= |u_m| \text{~for~}m=1,\dots,M, \label{eq:ym}
\end{align}

Henceforth, we assume additive white circular-Gaussian noise (AWGN) $w_m\sim\mc{N}(0,\nu^w)$.
Thus, if we condition on $z_m$, then $u_m$ is circular Gaussian with mean $z_m$ and variance $\nu^w$, and $y_m$ is Rician with pdf \cite{Rice:BSTJ:45}
\begin{align}
  \lefteqn{ p_{Y|Z}(y_m|z_m;\nu^w) }\nonumber\\
  &= \frac{2y_m}{\nu^w}
     \exp\Big(-\frac{y_m^2+|z_m|^2}{\nu^w}\Big)
     I_0\Big(\frac{2y_m|z_m|}{\nu^w}\Big)
     1_{y_m\geq 0} , \label{eq:pY|Z}
\end{align}
where $I_0(\cdot)$ is the $0^{th}$-order modified Bessel function of the first kind.

The functions $g\out(\cdot,\cdot)$ and $g\out'(\cdot,\cdot)$ defined in lines (D1)-(D3) of \tabref{gamp} can be computed using the expressions
\begin{align}
\lefteqn{ 
\E_{Z|Y,P}\{ Z|y_m,\hat{p}_m;\nu^p_m\} 
}\nonumber\\
&= \frac{\int_{\Complex} z\, p_{Y|Z}(y_m|z;\nu^w) \mc{N}(z;\hat{p}_m,\nu^p_m) dz}
	{\int_{\Complex} p_{Y|Z}(y_m|z';\nu^w) \mc{N}(z';\hat{p}_m,\nu^p_m) dz'} \\
&= \left( \frac{y_m}{1+\nu^w/\nu^p_m} R_0(\varrho_m)
        + \frac{|\hat{p}_m|}{\nu^p_m/\nu^w+1} \right) \frac{\hat{p}_m}{|\hat{p}_m|}
	\label{eq:Ez|yp}
\end{align}
and 
\begin{align}
\lefteqn{ 
\var_{Z|Y,P}\{ Z|y_m,\hat{p}_m;\nu^p_m\} 
}\nonumber\\
&= \frac{\int_{\Complex} |z|^2\, p_{Y|Z}(y_m|z;\nu^w) \mc{N}(z;\hat{p}_m,\nu^p_m) dz}
	{\int_{\Complex} p_{Y|Z}(y_m|z';\nu^w) \mc{N}(z';\hat{p}_m,\nu^p_m) dz'} 
\nonumber\\&\quad 
- |\E_{Z|Y,P}\{ Z|y_m,\hat{p}_m;\nu^p_m\}|^2 \\
&= \frac{y_m^2}{(1+\nu^w/\nu^p_m)^2}+\frac{|\hat{p}_m|^2}{(\nu^p_m/\nu^w+1)^2}
	+ \frac{1+\varrho_m R_0(\varrho_m)}{1/\nu^w+1/\nu^p_m} 
\nonumber\\&\quad 
- |\E_{Z|Y,P}\{ Z|y_m,\hat{p}_m;\nu^p_m\}|^2  ,
	\label{eq:Vz|yp}
\end{align}
where
\begin{align}
R_0(\varrho_m)\defn \frac{I_1(\varrho_m)}{I_0(\varrho_m)} 
\text{~and~}
\varrho_m \defn \frac{2y_m\,|\hat{p}_m|}{\nu^w+\nu^p_m} ,
\label{eq:R0}
\end{align}
as shown in \appref{gout}. 

\textb{
Whereas the above assumes that AWGN is added prior to the intensity step \eqref{ym}, it is also possible to consider post-intensity noise models, i.e.,
\begin{align}
y_m
&= q(|z_m|) + w_m ,
\label{eq:ym2}
\end{align}
where common examples of $q(\cdot)$ include $q(|z|)=|z|$ and $q(|z|)=|z|^2$
(see, e.g., \cite{Candes:CPAM:13}) 
and where $w_m \sim p_W$ for a specified noise distribution $p_W$.
The likelihood would then become
\begin{align}
p_{Y|Z}(y_m|z_m)
&= p_W(y_m-q(|z_m|)) ,
\label{eq:pY|Z2}
\end{align}
and the functions $g\out(\cdot,\cdot)$ and $g\out'(\cdot,\cdot)$ defined in lines (D1)-(D3) of \tabref{gamp} would be computed as described in \appref{gout2}.
Note that, to assign zero likelihood to negative intensity measurements $y_m$, the assumed noise density $p_W(w)$ must have zero measure on the negative reals.
}

\subsection{EM update of the noise variance} \label{sec:EM}

\textb{%
Until now we have treated the noise variance $\nu^w$ as a known parameter.  
In practice, however, $\nu^w$ may be unknown, in which case it is not clear what value to use in \eqref{Ez|yp} and \eqref{Vz|yp}.
To address this problem, we now describe how $\nu^w$ can be learned using an expectation-maximization (EM) \cite{Dempster:JRSS:77} procedure. 
The methodology is similar to that proposed in \cite{Krzakala:JSM:12}
for the case of a Gaussian $p_{Y|Z}(y_m|\cdot)$, but is more involved due to the fact that the $p_{Y|Z}(y_m|\cdot)$ used for phase-retrieval (recall \eqref{pY|Z}) is non-Gaussian.
}

\textb{%
Choosing $\vec{x}$ as the hidden data, the standard form of the $i$th EM update is \cite{Dempster:JRSS:77}
\begin{align}
\hat{\nu^w}[i\!+\!1]
&= \arg\max_{\nu^w > 0} \E\big\{\ln p(\vec{y},\vec{x};\nu^w) 
	\big| \vec{y}; \hat{\nu^w}[i] \big\} ,
	\label{eq:EM0}
\end{align}
where square brackets are used to distinguish EM iterations from GAMP iterations (recall \tabref{gamp}).
Because the true posterior pdf $p(\vec{x}|\vec{y})$ needed for \eqref{EM0} is generally NP-hard to compute \cite{Cooper:AI:90},
\appref{EM} describes an approximate EM update of the form
\begin{align}
\hat{\nu^w}(t\!+\!1)
&= \arg\min_{\nu^w > 0} \tilde{J}\big(\nu^w;\hvec{r}(t),\vec{\nu}^r(t),\hat{\nu^w}(t)\big) ,
\label{eq:EM}
\end{align}
which performs one EM iteration $i$ for every GAMP iteration $t$, allowing us to state the EM update \eqref{EM} using GAMP iterations.
In \eqref{EM}, $\tilde{J}$ is a certain Bethe free entropy and $(\hvec{r}(t),\vec{\nu}^r(t))$ are the results of lines (R5)-(R6) in \tabref{gamp} when GAMP is run under the noise variance $\hat{\nu^w}(t)$. 
(See \appref{EM} for details.)
}

\subsection{Signal prior distribution} \label{sec:prior}

GAMP offers great flexibility with respect to the choice of prior distribution on the signal vector $\vec{x}$.
In this work, we focus on separable priors, which have the form $p(\vec{x})=\prod_{n=1}^N p_{X_n}\!(x_n)$ with arbitrary $p_{X_n}\!(\cdot)$ (recalling \eqref{decoupled}), but we note that various forms of non-separable priors can be supported using the ``turbo GAMP'' formulation proposed in \cite{Schniter:CISS:10} or the ``analysis GAMP'' formulation proposed in \cite{Borgerding:13}.

For separable priors, $p_{X_n}\!(\cdot)$ should be chosen to reflect whatever form of probabilistic structure is known about coefficient $x_n$.
For example, if $\vec{x}\in\Complex^N$ is known to be $K$-sparse, but nothing is know about the support, then it is typical to choose the Bernoulli-Gaussian (BG) model
\begin{align}
p_{X_n}\!(x_n) 
&= (1-\lambda)\delta(x_n) + \lambda \mc{N}(x_n;0,\varphi) , \label{eq:BG}
\end{align}
with sparsity rate $\lambda=\frac{K}{N}$ and non-zero-coefficient variance $\varphi$ that, if unknown, can be estimated from the observations via \cite[eqn.\ (71)]{Vila:TSP:13}
\begin{align}
\varphi
&= \frac{\|\vec{y}\|_2^2- M\nu^w}{\lambda \|\vec{A}\|_F^2} ,
\end{align}
where $\|\cdot\|_F$ denotes the Frobenius norm. 
For this BG prior, expressions for the thresholding functions $g\inp(\cdot,\cdot)$ and $g\inp'(\cdot,\cdot)$ defined in lines (D5)-(D6) of \tabref{gamp} were given in \cite{Schniter:CISS:10}.
When the sparsity rate $\lambda$ in \eqref{BG} is unknown, it can be learned using the EM-BG procedure described in \cite{Vila:TSP:13}.
In most cases, improved performance is obtained when a Gaussian mixture (GM) pdf is used in place of the Gaussian pdf in \eqref{BG} \cite{Vila:TSP:13}.

Various extensions of the above are possible.
For example, 
when all coefficients $x_n$ are known to be real-valued or positive, the circular-Gaussian pdf in \eqref{BG} should be replaced by a real-Gaussian or truncated-Gaussian pdf, respectively, or even a truncated-GM \cite{Vila:CAMSAP:13}.
Furthermore, when certain coefficient subsets are known to be more or less sparse than others, a non-uniform sparsity \cite{Som:NAECON:10} rate $\lambda_n$ can be used in \eqref{BG}.

\subsection{GAMP normalization and damping}  \label{sec:step}

To increase the numerical robustness of GAMP, \textb{we propose to normalize} certain internal GAMP variables. 
To do this, we define $\alpha(t)\defn\frac{1}{M}\sum_{m=1}^M \nu_m^p(t)$ (which tends to grow very small with $t$ at high SNR), normalize both $\hat{s}_m(t)$ and $\nu^s_m(t)$ (which tend to grow very large) by $1/\alpha(t)$, and normalize $\nu^r_n(t)$ (which tends to grow very small) by $\alpha(t)$.
\textb{This prevents the normalized variables $\underline{\hat{s}}_m$, $\underline{\nu}^s_m(t)$, and $\underline{\nu}^r_n(t)$ from growing very large and causing numerical precision issues in Matlab.
We note that, under infinite precision, these normalizations would cancel each other out and have absolutely no effect.
The resulting normalized GAMP iterations are shown in \tabref{step}.} 

To reduce the chance of GAMP divergence, 
\textb{we propose to ``damp'' certain variable updates.  
Damping is a technique commonly used in loopy belief propagation (see, e.g., \cite{Heskes:NIPS:02}) to reduce the chance of divergence, although at the cost of convergence speed.
For GAMP, it was established in \cite{Rangan:ISIT:14} that damping is both necessary and sufficient to guarantee \emph{global} convergence under arbitrary $\vec{A}$ in the case of Gaussian $p_{X_n}(\cdot)$ and $p_{Y|Z}(y_m|\cdot)$. 
Similarly, \cite{Rangan:ISIT:14} established that damping is both necessary and sufficient to guarantee the \emph{local} convergence of GAMP under arbitrary $\vec{A}$ in the case of strictly log-concave $p_{X_n}(\cdot)$ and $p_{Y|Z}(y_m|\cdot)$.
For general $p_{X_n}(\cdot)$ and $p_{Y|Z}(y_m|\cdot)$, theory is (to the authors' knowledge) lacking, but empirical results (see, e.g., \cite{Vila:ICASSP:15}) suggest that the use of damping in GAMP can be very effective.
\tabref{step} presents a version of damped GAMP that uses a common damping parameter $\beta\in(0,1]$ throughout the algorithm: when $\beta=1$, the algorithm reduces to the original GAMP algorithm, but when $\beta<1$, the updates in lines (S1), (S4), (S5), and (S7) are slowed.
Our numerical experiments suggest that}
the value $\beta=0.25$ works well for phase retrieval.
One consequence of the proposed damping implementation is the existence of additional state variables like $\overline{x}_n(t)$.
To avoid the need to initialize these variables, we use $\beta=1$ during the first iteration.
We note that the damping modification described here is the one included in the public domain GAMPmatlab implementation,\footnote{\url{http://sourceforge.net/projects/gampmatlab/}} which differs slightly from the one described in \cite{Rangan:ISIT:14}.

\putTable{step}{GAMP steps with variance normalization $\alpha(t)$ and damping parameter $\beta\in(0,1]$.  
}{\footnotesize
\begin{equation*}
\begin{array}{|lr@{}c@{}l@{}r|}\hline
  \multicolumn{5}{|l|}{\texttt{for $t\!=\!1,2,3,\dots,T_{\max}$}}\\
  &\forall m:
   \nu^p_m(t)
   &=& \textstyle \beta \sum_{n=1}^{N} |a_{mn}|^2 \nu^x_n(t) + (1-\beta) \nu^p_m(t\!-\!1)& \text{(S1)}\\
  &\alpha(t)
   &=& \textstyle \frac{1}{M} \sum_{m=1}^M \nu^p_m(t) & \text{(S2)}\\
  &\forall m:
   \hat{p}_m(t)
   &=& \textstyle \sum_{n=1}^{N} \!a_{mn} \hat{x}_n(t) - \frac{\nu^p_m(t)}{\alpha(t)} \,\underline{\hat{s}}_m(t\!-\!1)& \text{(S3)}\\
  &\forall m:
   \underline{\hat{s}}_m(t)
   &=& \beta \alpha(t) g\out(\hat{p}_m(t),\nu^p_m(t)) + (1\!-\!\beta) \underline{\hat{s}}_m(t\!-\!1)& \text{(S4)}\\
  &\forall m:
   \underline{\nu}^s_m(t)
   &=& -\beta \alpha(t) g'\out(\hat{p}_m(t),\nu^p_m(t)) \!+\! (1\!-\!\beta) \underline{\nu}^s_m(t\!-\!1)& \text{(S5)}\\
  &\forall n:
   \underline{\nu}^r_n(t)
   &=& \textstyle \big(\sum_{m=1}^{M} |a_{mn}|^2 \underline{\nu}^s_m(t) 
        \big)^{-1} & \text{(S6)}\\
  &\forall n:
   \overline{x}_n(t)
   &=& \textstyle \beta \hat{x}_n(t) + (1\!-\!\beta) \overline{x}_n(t\!-\!1) & \text{(S7)}\\
  &\forall n:
   \hat{r}_n(t)
   &=& \textstyle \overline{x}_n(t)+ \underline{\nu}^r_n(t) \sum_{m=1}^{M} \!a_{mn}^*
        \underline{\hat{s}}_m(t)  & \text{(S8)}\\
  &\forall n\!:
   \nu^x_n(t\!+\!1)
   &=& \alpha(t) \underline{\nu}^r_n(t) g'\inp\big(\hat{r}_n(t),\alpha(t)\underline{\nu}^r_n(t)\big) & \text{(S9)}\\
  &\forall n\!:
   \hat{x}_n(t\!+\!1)
   &=& g\inp\big(\hat{r}_n(t),\alpha(t)\underline{\nu}^r_n(t)\big) & \text{(S10)}\\
  \multicolumn{5}{|l|}{\texttt{end}}\\\hline
\end{array}
\end{equation*}
}

\subsection{Avoiding bad local minima} \label{sec:init}

As is well known \textb{\cite{Moravec:SPIE:07,Shechtman:TSP:14,Mukherjee:TSP:14,Netrapalli:NIPS:13}}, the compressive phase retrieval problem is plagued by bad local minima.
We now propose methods to \textb{randomly} initialize and restart PR-GAMP that aim to avoid these bad local minima.
\textb{Our empirical experience (see \secref{num}) suggests that the existence of bad local minima is a more serious issue with Fourier $\vec{A}$ than with randomized (e.g., i.i.d Gaussian or masked-Fourier) $\vec{A}$.}

\subsubsection{GAMP initialization}
The GAMP algorithm in \tabref{gamp} requires an initialization of the signal coefficient estimates $\{\hat{x}_n(1)\}_{n=1}^N$, their variances $\{\nu^x_n(1)\}_{n=1}^N$, and the state variables $\{\hat{s}_m(0)\}_{m=1}^M$ (which can be interpreted as Lagrange multipliers \cite{Rangan:ISIT:14}).
\textb{The standard procedure outlined in \cite{Rangan:ISIT:11} uses the fixed initialization $\hat{x}_n(1)=\E\{X_n\}$, $\nu^x_n(1)=\var\{X_n\}$, $\hat{s}_m(0)=0$.
But, from this fixed initialization, GAMP may converge to a bad local minimum.
To allow the possibility of avoiding this bad local minima, we propose to \emph{randomly} initialize and restart GAMP multiple times if needed.
For the random initializations, 
we propose to draw each $\hat{x}_n(1)$ as an independent realization of the random variable $X_n$.
This way, the empirical mean of $\{\hat{x}_n(1)\}_{n=1}^N$ matches that of the standard initialization from \cite{Rangan:ISIT:11}. 
Likewise, we propose to initialize $\nu^x_n(1)$, for all $n$, at the empirical variance of $\{\hat{x}_n(1)\}_{n=1}^N$. 
}

\subsubsection{EM initialization}
For the EM algorithm described in \secref{EM}, we must choose the initial noise-variance estimate $\hat{\nu^w}(1)$.
Even when accurate knowledge of $\nu^w$ is available, our \textb{numerical experience leads us to believe that} setting $\hat{\nu^w}(1)$ at a relatively large value can help to avoid bad local minima.
In particular, our \textb{empirical experience} leads us to suggest setting $\hat{\nu^w}(1)$ in correspondence with an initial SNR estimate of $10$, i.e.,
$\hat{\nu^w}(1) = \frac{\|\vec{y}\|_2^2}{M(\SNR\init+1)}$ with $\SNR\init=10$.

\subsubsection{Multiple restarts}
To further facilitate the avoidance of bad local minima, we propose to run multiple attempts of EM-GAMP, each using a different random GAMP initialization (constructed as above).
The attempt that yields the lowest normalized residual ($\NRES\defn \|\vec{y}-|\vec{A}\hvec{x}|\|_2^2/\|\vec{y}\|_2^2$) is then selected as the algorithm output.
The efficacy of multiple attempts is numerically investigated in \secref{num}.

Furthermore, to avoid unnecessary restarts, we allow the algorithm to be stopped as soon as the $\NRES$ drops below a user-defined stopping tolerance of $\NRES\suff$.
When the true SNR is known, we suggest setting $\NRES\suff\dB=-(\SNR\true\dB+2)$.

\subsubsection{Algorithm summary}
The PR-GAMP algorithm is summarized in \tabref{prgamp}, where 
$A_{\max}$ controls the number of attempts,
$\SNR\init$ controls the initial SNR, and
$\NRES\suff$ controls the stopping tolerance.

\putTable{prgamp}{The proposed PR-GAMP algorithm with $A_{\max}$ attempts, SNR initialization \textrm{$\SNR\init$}, and stopping residual $\NRES\suff$.}{\footnotesize
\textb{
\begin{equation*}
\begin{array}{|lll|}\hline
  \multicolumn{3}{|l|}{\texttt{input 
  	$\vec{y},\vec{A},
	 \{p_{X_n}\!(\cdot)\}_{n=1}^N,
	 \SNR\init, \NRES\suff, A_{\max}, T_{\max}$
	 }}\\[0.5mm]
  \multicolumn{3}{|l|}{
  	 \hat{\nu^w}(1)=\displaystyle \frac{\|\vec{y}\|_2^2}{M(\SNR\init+1)}
	 }\\[1mm]
  \multicolumn{3}{|l|}{
  	 \forall m: \hat{s}_m(0)=0
	 }\\
  \multicolumn{3}{|l|}{
  	 \NRES\best=\infty
	 }\\[1mm]
  \multicolumn{3}{|l|}{\texttt{for $a=1,2,3,\dots,A_{\max}$,}}\\
  \quad&\multicolumn{2}{l|}{
  	\texttt{draw random $\hvec{x}(1)$} 
  } \\
  \quad&\multicolumn{2}{l|}{
  	\forall n: \nu^x_n(1) = \|\hvec{x}(1)-\E\{\vec{X}\}\|_2^2/N
  } \\
  \quad&\multicolumn{2}{l|}{
  	\texttt{for $t=1,2,3,\dots,T_{\max}$}
  } \\
  \quad&\quad& 
  		\big(\hvec{x}(t\!+\!1),\vec{\nu}^x(t\!+\!1),\hvec{s}(t),\hvec{r}(t),\vec{\nu}^r(t)\big)\\
  \quad&\quad&
		= \textsf{GAMP}\big(
		  \vec{A},
		  \{p_{X_n}\!(\cdot)\}_{n=1}^N,
		  \{p_{Y|Z}(y_m|\cdot;\hat{\nu^w}(t))\}_{m=1}^M, \\
  \quad&\quad&\hspace{13mm}
  		  \hvec{x}(t),\vec{\nu}^x(t),\hvec{s}(t\!-\!1)
		\big)\\
  \quad&\quad& 
  		\hat{\nu^w}(t\!+\!1)
                = \arg\min_{\nu^w > 0} \tilde{J}\big(\nu^w;\hvec{r}(t),\vec{\nu}^r(t),\hat{\nu^w}(t)\big) \\
  \quad&\multicolumn{2}{l|}{
  	\texttt{end}
  }\\
  \quad&\multicolumn{2}{l|}{
  	\NRES=\|\vec{y}-|\vec{A}\hvec{x}(T_{\max}\!+\!1)|\|_2^2/\|\vec{y}\|_2^2
  } \\
  \quad&\multicolumn{2}{l|}{
  	\texttt{if $\NRES < \NRES\best$}
  }\\
  \quad&\quad& 
  		\hvec{x}\best = \hvec{x}(T_{\max}\!+\!1) \\
  \quad&\quad& 
  		\NRES\best = \NRES \\
  \quad&\multicolumn{2}{l|}{
  	\texttt{end}
  }\\
  \quad&\multicolumn{2}{l|}{
  	\texttt{if $\NRES < \NRES\suff$}
  }\\
  \quad&\quad& 
  		\texttt{stop}\\
  \quad&\multicolumn{2}{l|}{
  	\texttt{end}
  }\\
  \multicolumn{3}{|l|}{\texttt{end}}\\[1mm]
  \multicolumn{3}{|l|}{\texttt{output 
  	$\hvec{x}\best$
	}}\\[1mm]
  \hline
\end{array}
\end{equation*}
}
}

\section{Numerical Results}	\label{sec:num}

In this section we numerically investigate the performance of PR-GAMP\footnote{PR-GAMP is part of the GAMPmatlab package at \url{http://sourceforge.net/projects/gampmatlab/}.} under various scenarios and in comparison to several existing algorithms:
Compressive Phase Retrieval via Lifting (CPRL) \cite{Ohlsson:NIPS:12},
GrEedy Sparse PhAse Retrieval (GESPAR) from \cite{Shechtman:TSP:14},
and the sparse Fienup technique from \cite{Mukherjee:TSP:14},
As a benchmark, we also compare to ``phase oracle'' (PO) GAMP,
i.e., GAMP operating on the magnitude-and-phase measurements $\vec{u}=\vec{Ax}+\vec{w}$ rather than on the intensity measurements $\vec{y}=|\vec{u}|$. 

Unless otherwise noted, we generated random realizations the true signal vector $\vec{x}$ as $K$-sparse length-$N$ with support chosen uniformly at random and with nonzero coefficients drawn i.i.d zero-mean circular-Gaussian.
Then, for a given matrix $\vec{A}$, we generated $M$ noisy intensity measurements $\vec{y}=|\vec{Ax}+\vec{w}|$, where $\vec{w}$ was i.i.d circular-Gaussian with variance selected to achieve a target signal-to-noise ratio of $\SNR\defn\norm{\vec{Ax}}_2^2/\E\{\norm{\vec{w}}_2^2\}$.
Finally, each algorithm computed an estimate $\hvec{x}$ from $(\vec{y},\vec{A})$ in an attempt to best match $\vec{x}$ up to \textb{the inherent level of ambiguity.
We recall that, for any $\vec{A}$, the magnitude $|\vec{Ax}|$ is invariant to global phase rotations in $\vec{x}$.
For Fourier $\vec{A}$ and real-valued $\vec{x}$, the magnitudes of $\vec{Ax}$ are also invariant to flips and circular shifts of $\vec{x}$.}
Performance was then assessed using normalized mean-squared error on the disambiguated estimate:
\begin{equation}
  \NMSE(\hvec{x}) 
  \defn \min_{\vec{\Theta}} \frac{\norm{\vec{x}-\textsf{disambig}(\hvec{x},\vec{\Theta})}_2^2}{\norm{\vec{x}}_2^2} ,
\end{equation}
where $\vec{\Theta}$ are the ambiguity parameters.
When computing empirical phase-transition curves, we defined a ``successful'' recovery as one that produced $\NMSE < 10^{-6}$.

\subsection{Empirical phase transitions: i.i.d Gaussian $\vec{A}$}	\label{sec:ptc}

First we investigated the phase-transition performance of PR-GAMP with i.i.d circular-Gaussian sensing matrices $\vec{A}$. 
\Figref{pr10_success_iid} plots the empirical success rate (averaged over $100$ independent problem realizations) as a function of signal sparsity $K$ and measurement length $M$ for a fixed signal length of $N=512$.
Here we used $\SNR=100$~dB, which makes the observations essentially ``noiseless,''
and we allowed PR-GAMP up to $10$ attempts from random initializations (i.e., $A_{\max}=10$ in \tabref{prgamp}).
The figure shows a ``phase transition'' behavior that separates the $(K,M)$ plane into two regions: perfect recovery in the top-left and failure in the bottom-right.
Moreover, the figure \textb{suggests} that, \textb{to recover $K$-sparse Bernoulli-Gaussian signals with} $K\ll N$, approximately $M\geq 2K \log_2(N/K)$ intensity measurements suffice for PR-GAMP.

To investigate how well (versus how often) PR-GAMP recovers the signal, we plot the median $\NMSE$ \textb{achieved} over the same problem realizations in \figref{pr10_nmse_iid}.  
There we see that the signal estimates \textb{were} extremely accurate \textb{on the good side of the phase transition}.

To investigate the effect of number-of-attempts $A_{\max}$, we extracted the $50$\%-success contour (i.e., the phase-transition curve) from \figref{pr10_success_iid} and plotted it in \figref{prpo_ptc_iid}, along with the corresponding contours obtained under different choices of $A_{\max}$.
\Figref{prpo_ptc_iid} \textb{suggests} that, in the case of i.i.d $\vec{A}$, there is relatively little to gain from multiple restarts from random realizations. 
With Fourier $\vec{A}$, however, we will see in the sequel that multiple restarts are indeed important.

\Figref{prpo_ptc_iid} also plots the phase-transition curve of phase-oracle (PO)-GAMP calculated from the same problem realizations.
A comparison of the PO-GAMP phase transition to the PR-GAMP phase transition \textb{suggests} that PR-GAMP requires approximately $4\times$ the number of measurements as PO-GAMP, regardless of sparsity rate $K$, \textb{for Bernoulli-Gaussian signals}.
Remarkably, this ``$4\times$'' rule generalizes what is known about the recovery of \emph{non}-sparse signals in $\Complex^N$, where the ratio of (necessary and sufficient) magnitude-only to magnitude-and-phase measurements is also $4\times$ (as $N\rightarrow\infty$) \cite{Heinosaari:CMP:13,Bodmann:ACM:14}.

Overall, Figures~\ref{fig:pr10_success_iid}--\ref{fig:prpo_ptc_iid} demonstrate that PR-GAMP is indeed capable of \emph{compressive} phase retrieval, i.e., successful $\Complex^N$-signal recovery from $M\ll 4N$ intensity measurements, when the signal is sufficiently sparse.
Moreover, to our knowledge, these phase transitions are far better than those \textb{reported for existing} algorithms in the literature.

\putFrag{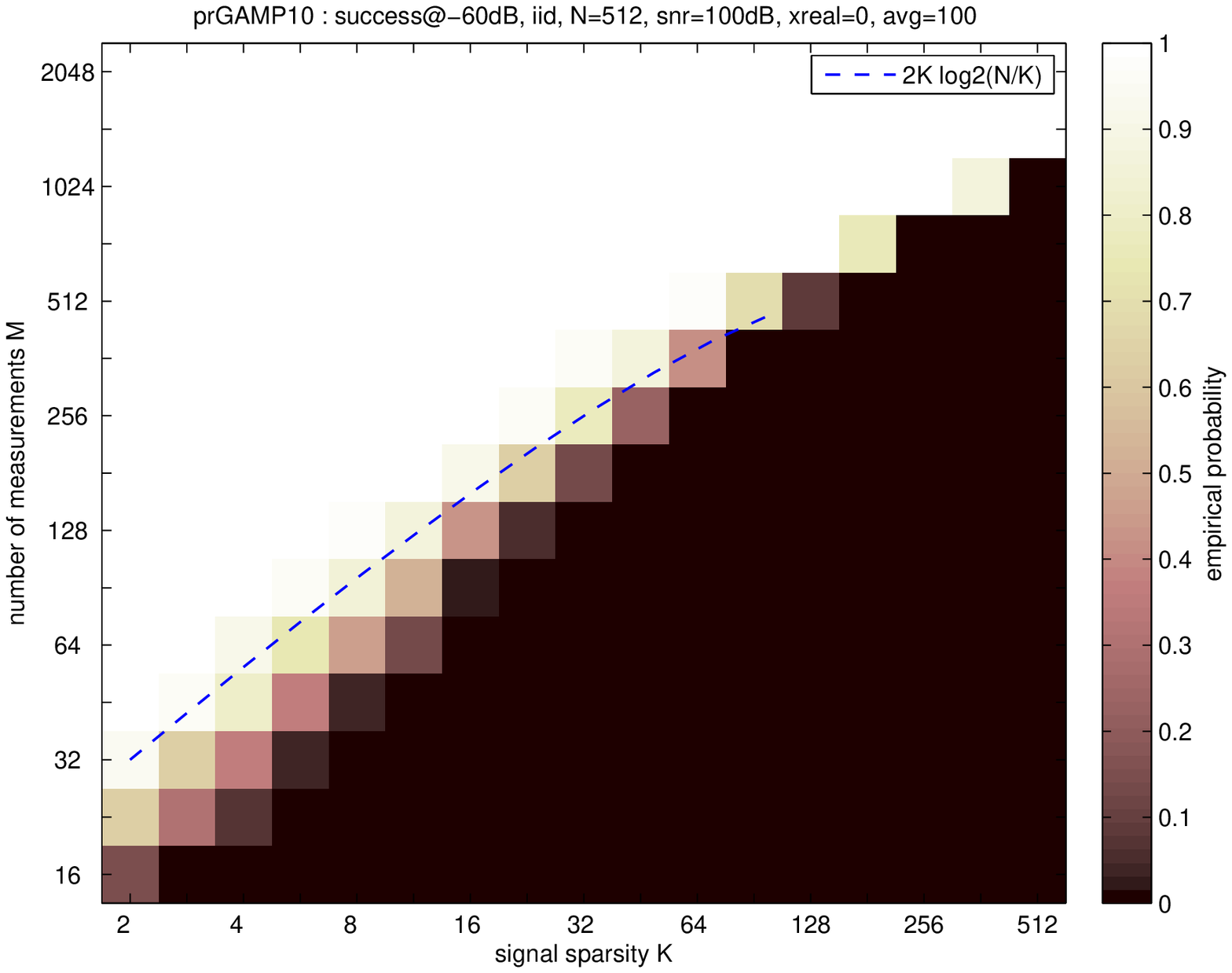}
	{Empirical probability of successful PR-GAMP recovery of an $N=512$-length signal, versus signal sparsity $K$ and number of intensity measurements $M$, using i.i.d Gaussian $\vec{A}$ at $\SNR\!=\!100$~dB.  Here, PR-GAMP was allowed up to $10$ attempts from different random initializations.}
	{\figsize}
	{\psfrag{signal sparsity K}[t][t][0.6]{\sf sparsity $K$} 
	 \psfrag{number of measurements M}[b][b][0.6]{\sf measurements $M$}
	 \psfrag{2K log2(N/K)}[lb][lB][0.45]{$2K\log_2(N/K)$}
	 \psfrag{prGAMP10 : success@-60dB, iid, N=512, snr=100dB, xreal=0, avg=100}{}
	 \psfrag{empirical probability}[t][t][0.6]{\sf empirical success rate}
	 }

\putFrag{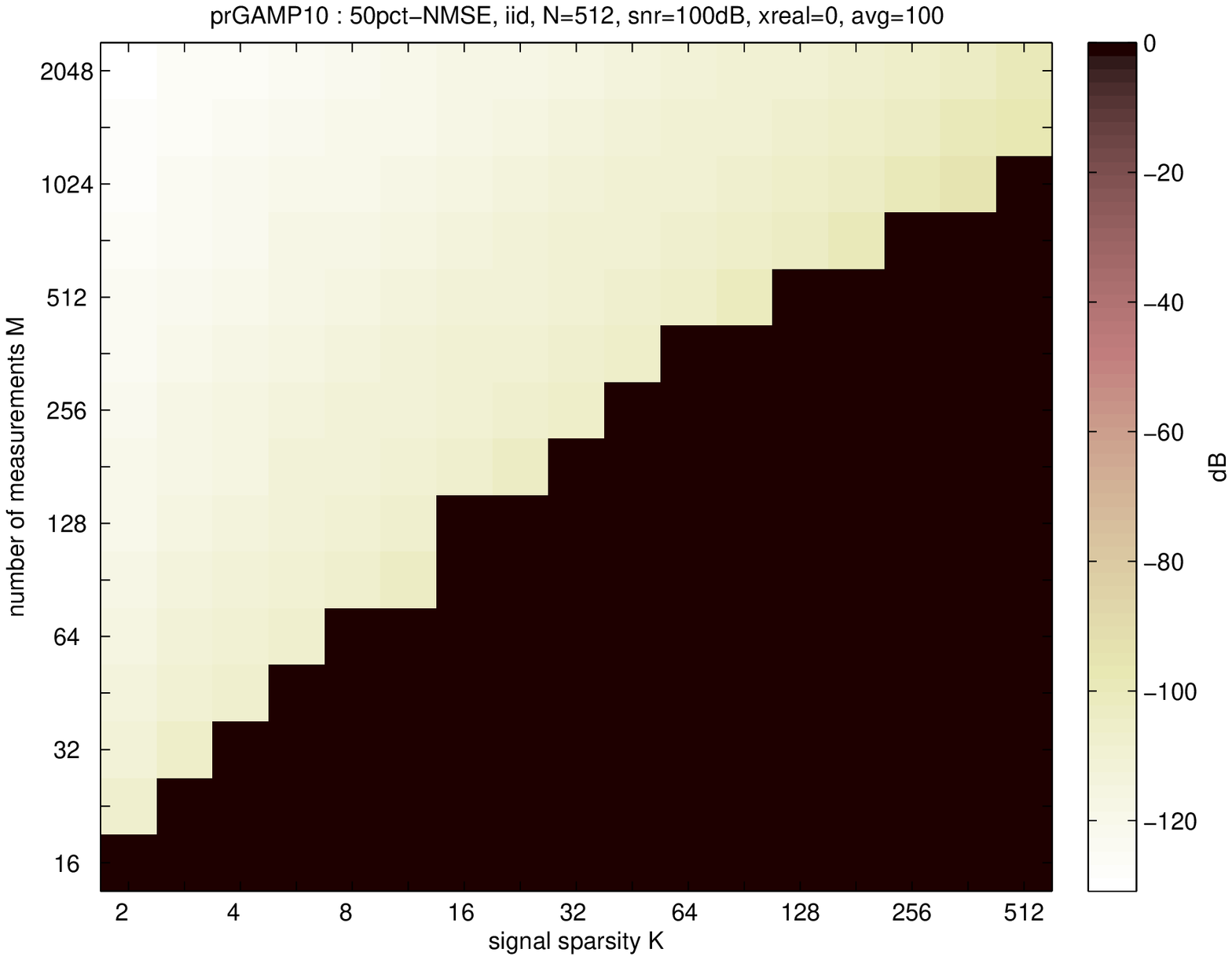}
	{Median $\NMSE$ for PR-GAMP recovery of an $N\!=\!512$-length signal, versus signal sparsity $K$ and number of intensity measurements $M$, using i.i.d Gaussian $\vec{A}$ at $\SNR\!=\!100$~dB.  Here, PR-GAMP was allowed up to $10$ attempts from different random initializations.}
	{\figsize}
	{\psfrag{signal sparsity K}[t][t][0.6]{\sf sparsity $K$} 
	 \psfrag{number of measurements M}[b][b][0.6]{\sf measurements $M$}
	 \psfrag{dB}[t][t][0.6]{\sf dB}
	 \psfrag{prGAMP10 : 50pct-NMSE, iid, N=512, snr=100dB, xreal=0, avg=100}{}}

\putFrag{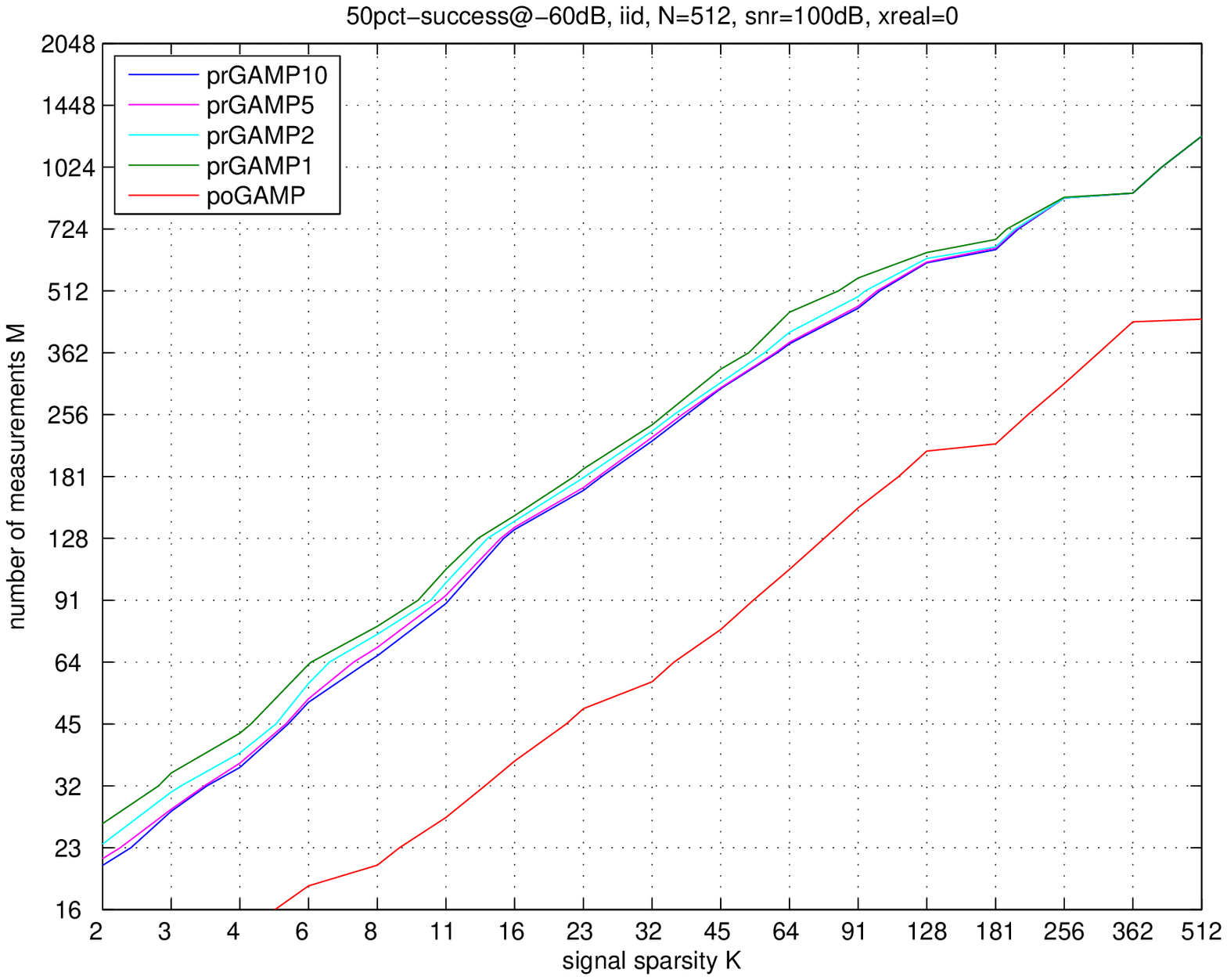}
	{50\%-success contours for PR-GAMP and phase-oracle GAMP recovery of an $N\!=\!512$-length signal, versus signal sparsity $K$ and number of intensity measurements $M$, using i.i.d Gaussian $\vec{A}$ at $\SNR\!=\!100$~dB.  PR-GAMP-$A_{\max}$ denotes PR-GAMP under a maximum of $A_{\max}$ attempts.}
	{3.2}
	{\psfrag{signal sparsity K}[t][t][0.6]{\sf sparsity $K$} 
	 \psfrag{number of measurements M}[b][b][0.6]{\sf measurements $M$} 
	 \psfrag{prGAMP1}[l][l][0.4]{\sf PR-GAMP-1} 
	 \psfrag{prGAMP2}[l][l][0.4]{\sf PR-GAMP-2} 
	 \psfrag{prGAMP5}[l][l][0.4]{\sf PR-GAMP-5} 
	 \psfrag{prGAMP10}[l][l][0.4]{\sf PR-GAMP-10} 
	 \psfrag{poGAMP}[l][l][0.4]{\sf PO-GAMP}
	 \psfrag{50pct-success@-60dB, iid, N=512, snr=100dB, xreal=0}{}}

\subsection{Robustness to noise}		\label{sec:noise}

We now demonstrate the robustness of PR-GAMP to non-trivial levels of additive white circular-Gaussian noise $\vec{w}$ in the $M$ intensity measurements $\vec{y} = |\vec{Ax}+\vec{w}|$.
As before, we use $N=512$-length $K$-sparse \textb{Bernoulli-Gaussian} signals and i.i.d Gaussian $\vec{A}$, but now we focus on \textb{sparsity $K=4$ and number of measurements $M\in\{64,128,256\}$. 
We note that these $(K,M)$ pairs are all on the good side of the phase-transition in \figref{pr10_success_iid}, although $(K,M)=(4,64)$ is near the boundary.}
\Figref{prpo_snr_iid} shows median $\NMSE$ performance over $200$ independent problem realizations as a function of $\SNR\defn\norm{\vec{Ax}}_2^2/\norm{\vec{w}}_2^2$.
There we see that, \textb{for most of the tested $(M,\SNR)$ pairs}, PR-GAMP performs only about $3$~dB worse than PO-GAMP. 
This $3$~dB gap can be explained by the fact that PO-GAMP is able to average the noise over twice as many real-valued measurements as PR-GAMP (i.e., $\{\real\{u_m\},\imag\{u_m\}\}_{m=1}^M$ versus $\{|u_m|\}_{m=1}^M$).
\textb{\Figref{prpo_snr_iid} shows that the performance gap grows beyond $3$~dB when both the $\SNR$ is very low and the measurements are very few.
But this may reflect a fundamental performance limitation rather than a weakness in PR-GAMP.}

\putFrag{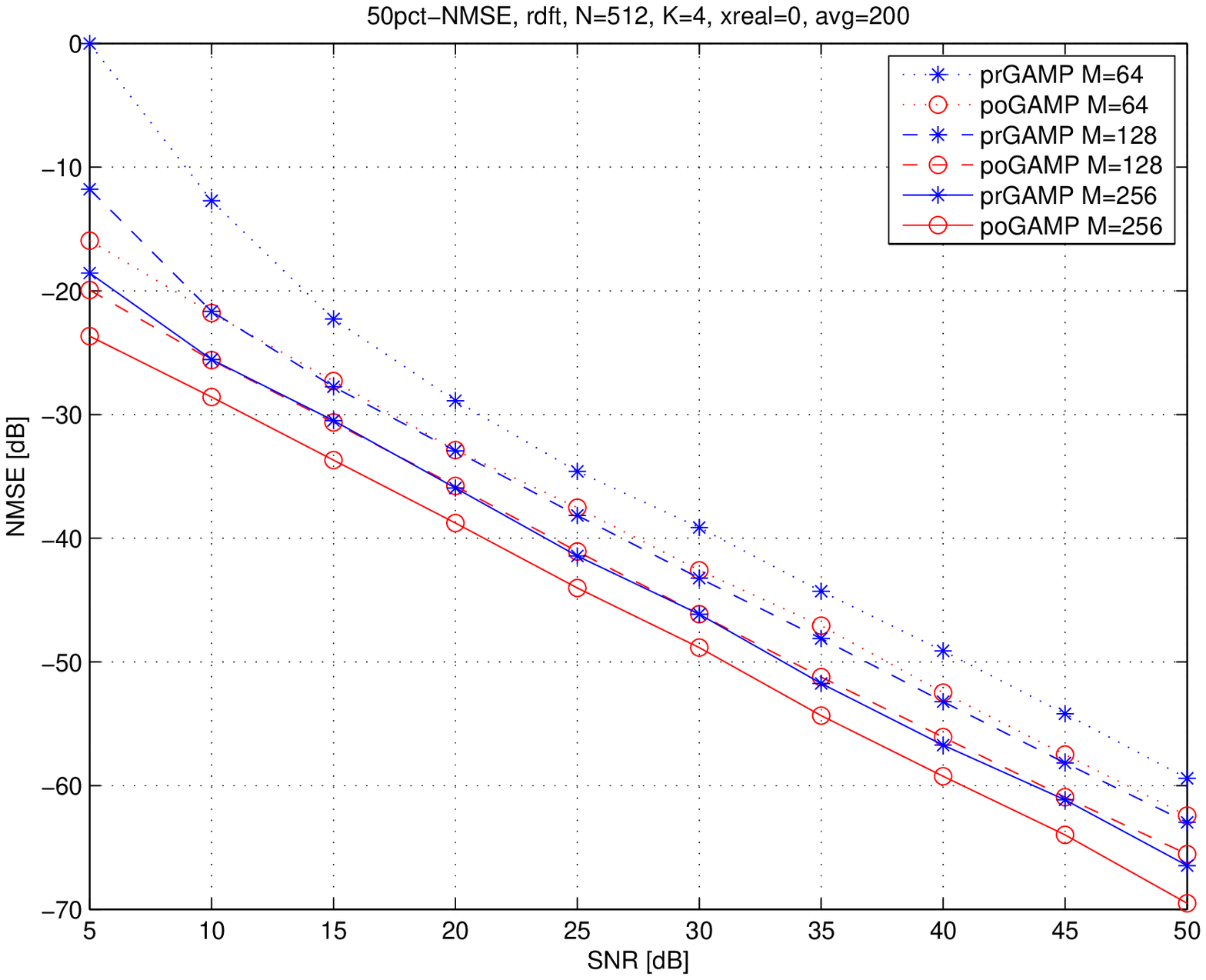}
	{Median $\NMSE$ for PR-GAMP and phase-oracle GAMP recovery of an $N\!=\!512$-length $K\!=\!4$-sparse signal, versus $\SNR$, \textb{from $M\!\in\!\{64,128,256\}$} measurements and i.i.d Gaussian $\vec{A}$.}
	{3.2}
	{\psfrag{NMSE [dB]}[b][b][0.6]{\sf NMSE in dB} 
	 \psfrag{SNR [dB]}[t][t][0.6]{\sf SNR in dB} 
	 \psfrag{50pct-NMSE, rdft, N=512, K=4, xreal=0, avg=200}{}
	 \psfrag{prGAMP M=64}[l][l][0.4]{\sf \!PR-GAMP, $M\!=\!64$}
	 \psfrag{poGAMP M=64}[l][l][0.4]{\sf \!PO-GAMP, $M\!=\!64$}
	 \psfrag{prGAMP M=128}[l][l][0.4]{\sf \!PR-GAMP, $M\!=\!128$}
	 \psfrag{poGAMP M=128}[l][l][0.4]{\sf \!PO-GAMP, $M\!=\!128$}
	 \psfrag{prGAMP M=256}[l][l][0.4]{\sf \!PR-GAMP, $M\!=\!256$}
	 \psfrag{poGAMP M=256}[l][l][0.4]{\sf \!PO-GAMP, $M\!=\!256$}
         }

\subsection{Comparison to CPRL}			\label{sec:cprl}
In this section, we present compare PR-GAMP to the state-of-the-art convex-relaxation approach to compressive phase retrieval, CPRL \cite{Ohlsson:NIPS:12}.
To implement CPRL, we used the authors' CVX-based matlab code\footnote{\url{http://users.isy.liu.se/rt/ohlsson/code/CPRL.zip}} under default algorithmic settings.
We also tried the authors' ADMM implementation, but found that it gave significantly worse performance.
As before, we examine the recovery of a $K$-sparse signal in $\Complex^N$ from $M$ intensity measurements $\vec{y}=|\vec{Ax}+\vec{w}|$, but now we use $\vec{A}=\vec{\Phi F}$ with i.i.d circular-Gaussian $\vec{\Phi}$ and discrete Fourier transform (DFT) $\vec{F}$, to be consistent with the setup assumed in \cite{Ohlsson:NIPS:12}.

\tabref{k1} shows empirical success\footnote{Since CPRL rarely gave $\NMSE<10^{-6}$, we reduced the definition of ``success'' to $\NMSE<10^{-4}$ for this subsection only.} rate and runtime (on a standard personal computer) for a problem with sparsity $K=1$, signal lengths $N\in\{32,48,64\}$, and compressive measurement lengths $M\in\{20,30,40\}$.
The table shows that, over $100$ problem realizations, both algorithms were $100$\% successful in recovering the signal at all tested combinations of $(M,N)$. 
But the table also shows that CPRL's runtime increased rapidly with the signal dimensions, whereas that of PR-GAMP remained orders-of-magnitude smaller and \textb{relatively} independent of $(M,N)$ over the tested range.\footnote{Although the complexity of GAMP is known to scale as $O(MN)$ for this type of $\vec{A}$, the values of $M$ and $N$ in \tabref{k1} and \tabref{k2} are too small for this scaling law to manifest. \textb{Instead, the runtime values in these tables are biased by the overhead computations associated with Matlab's object-oriented programming environment.}}

\tabref{k2} repeats the experiment carried out in \tabref{k1}, but at the sparsity $K=2$.
For this more difficult problem, the table shows that CPRL was much less successful at recovering the signal than PR-GAMP.
Meanwhile, the runtimes reported in \tabref{k2} again show that CPRL's complexity scaled rapidly with the problem dimensions, whereas GAMP's complexity stayed orders-of-magnitude smaller and \textb{relatively} constant over the tested problem dimensions.
In fact, the comparisons conducted in this section were restricted to very small problem dimensions precisely due to the poor complexity scaling of CPRL.

\putTable{k1}
{Empirical success rate and median runtime over $100$ problem realizations for several combinations of signal length $N$, measurement length $M$, and signal sparsity $K=1$.}
{\begin{tabular}{|@{\;}c@{\;}|@{\;}c@{\;}|@{\;}c@{\;}|@{\;}c@{\;}|}\hline
   & \scriptsize $(M,N)=(20,32)$ & \scriptsize $(M,N)=(30,48)$ & \scriptsize $(M,N)=(40,64)$ \\\hline
   \scriptsize CPRL & 1.00 (3.4 sec) & 1.00 (37 sec) & 1.00 (434 sec) \\
   \scriptsize PR-GAMP & \textb{1.00 (0.18 sec)} & \textb{1.00 (0.17 sec)} & \textb{1.00 (0.16 sec)} \\\hline
 \end{tabular}
}

\putTable{k2}
{Empirical success rate and median runtime over $100$ problem realizations for several combinations of signal length $N$, measurement length $M$, and signal sparsity $K=2$.}
{\begin{tabular}{|@{\;}c@{\;}|@{\;}c@{\;}|@{\;}c@{\;}|@{\;}c@{\;}|}\hline
   & \scriptsize $(M,N)=(20,32)$ & \scriptsize $(M,N)=(30,48)$ & \scriptsize $(M,N)=(40,64)$ \\\hline
   \scriptsize CPRL & 0.55 (4.1 sec) & 0.65 (42 sec) & 0.66 (496 sec) \\
   \scriptsize PR-GAMP & \textb{0.93 (0.25 sec)} & \textb{1.00 (0.21 sec)} & \textb{1.00 (0.19 sec)} \\\hline
 \end{tabular}
}

\subsection{Comparison to sparse-Fienup and GESPAR: Fourier $\vec{A}$}	\label{sec:gespar}

In this section, we compare PR-GAMP to the sparse-Fienup \cite{Mukherjee:TSP:14} and GESPAR\footnote{For GESPAR, we used the November 2013 version of the Matlab code provided by the authors at \url{https://sites.google.com/site/yoavshechtman/resources/software}.} \cite{Shechtman:TSP:14} algorithms. 
This comparison requires\footnote{The sparse Fienup from \cite{Mukherjee:TSP:14} requires $\vec{A}\herm\vec{A}$ to be a (scaled) identity matrix. Although GESPAR can in principle handle generic $\vec{A}$, the implementation provided by the authors is based on 1D and 2D Fourier $\vec{A}$ and is not easily modified.} that we restrict our attention to Fourier-based $\vec{A}$ and real-valued sparse vectors $\vec{x}$. 
For the experiments below, we generated realizations of $\vec{x}$ \textb{as described earlier}, but now with the non-zero elements drawn from a real-Gaussian distribution.
Also, we used $ITER=6400$ in GESPAR as recommended by the authors in \cite{Shechtman:TSP:14}, and we allowed sparse-Fienup $1000$ attempts from random initializations.

We first consider 2D Fourier $\vec{A}$, which is especially important for imaging applications. 
In particular, we repeat an experiment from \cite{Shechtman:TSP:14}, where the measurement and signal lengths were fixed at $M=N$ and the signal sparsity $K$ was varied.
For $N=1024$, \figref{success_2Dfft} shows the empirical success rate (over $200$ realizations) for PR-GAMP, GESPAR, and sparse Fienup.
Meanwhile, \figref{time_2Dfft} shows the corresponding median runtime for each algorithm, where all algorithms leveraged fast Fourier transform (FFT) implementations of $\vec{A}$.
From \figref{success_2Dfft}, we can see that PR-GAMP produced a significantly better phase-transition than GESPAR and sparse Fienup.
Meanwhile, from \figref{time_2Dfft} we see that, for the challenging case of $K\geq 40$, PR-GAMP-10 had uniformly better runtime \emph{and} success rate than GESPAR and sparse Fienup.

\putFrag{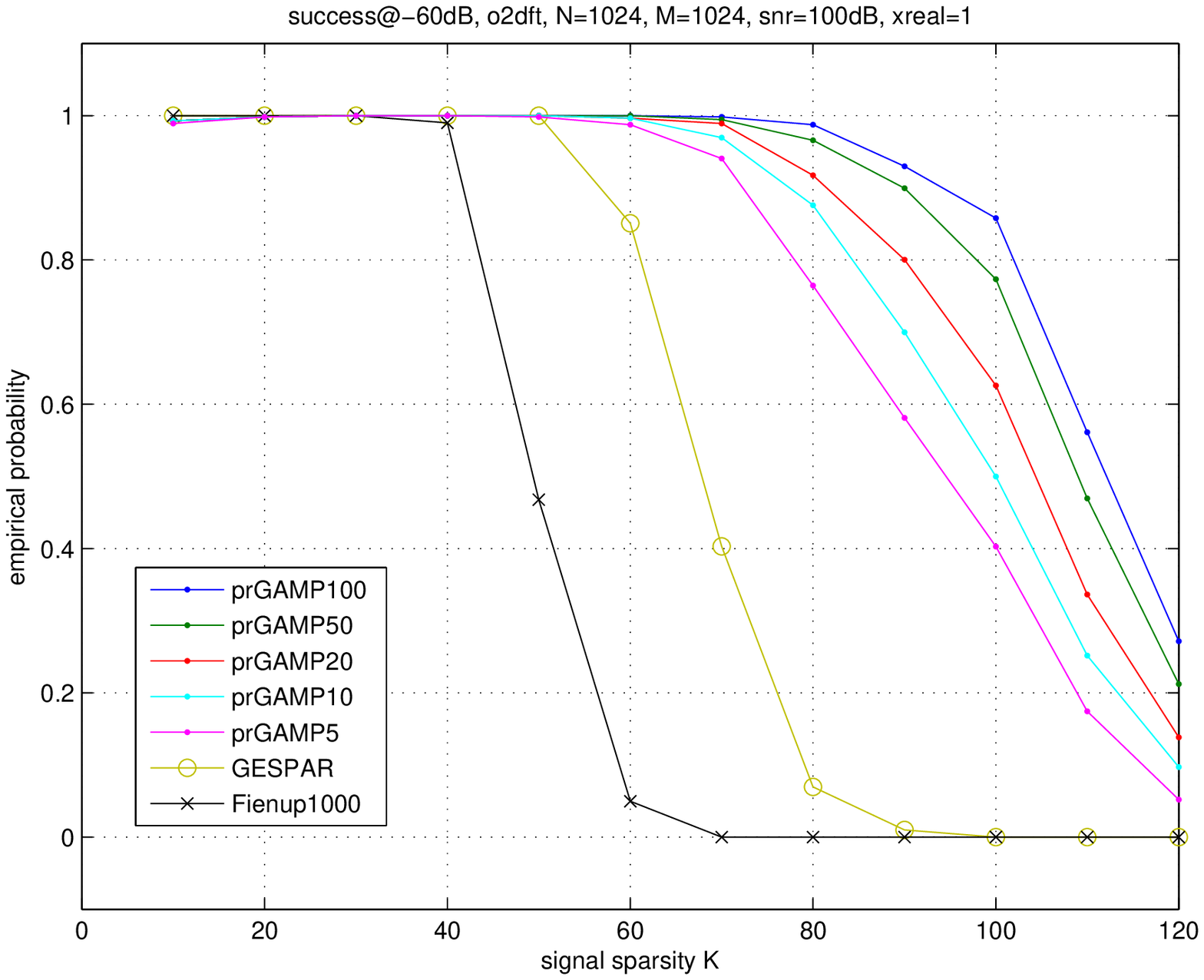}
	{Empirical success rate versus sparsity $K$ in the recovery of an $N\!=\!1024$-length real-valued signal from $M\!=\!1024$ 2D-Fourier intensities at $\SNR=100$dB. PR-GAMP-$A$ denotes PR-GAMP under a maximum of $A$ attempts.}
	{3.2}
	{\psfrag{signal sparsity K}[t][t][0.6]{\sf sparsity $K$} 
	 \psfrag{prGAMP200}[l][l][0.43]{\sf PR-GAMP-200} 
	 \psfrag{prGAMP100}[l][l][0.43]{\sf PR-GAMP-100} 
	 \psfrag{prGAMP50}[l][l][0.43]{\sf PR-GAMP-50} 
	 \psfrag{prGAMP20}[l][l][0.43]{\sf PR-GAMP-20} 
	 \psfrag{prGAMP10}[l][l][0.43]{\sf PR-GAMP-10} 
	 \psfrag{prGAMP5}[l][l][0.43]{\sf PR-GAMP-5} 
	 \psfrag{GESPAR}[l][l][0.43]{\sf GESPAR}
	 \psfrag{Fienup1000}[l][l][0.43]{\sf Fienup}
	 \psfrag{empirical probability}[b][b][0.6]{\sf empirical success rate}
	 \psfrag{success@-60dB, o2dft, N=1024, M=1024, snr=100dB, xreal=1}{}}

\putFrag{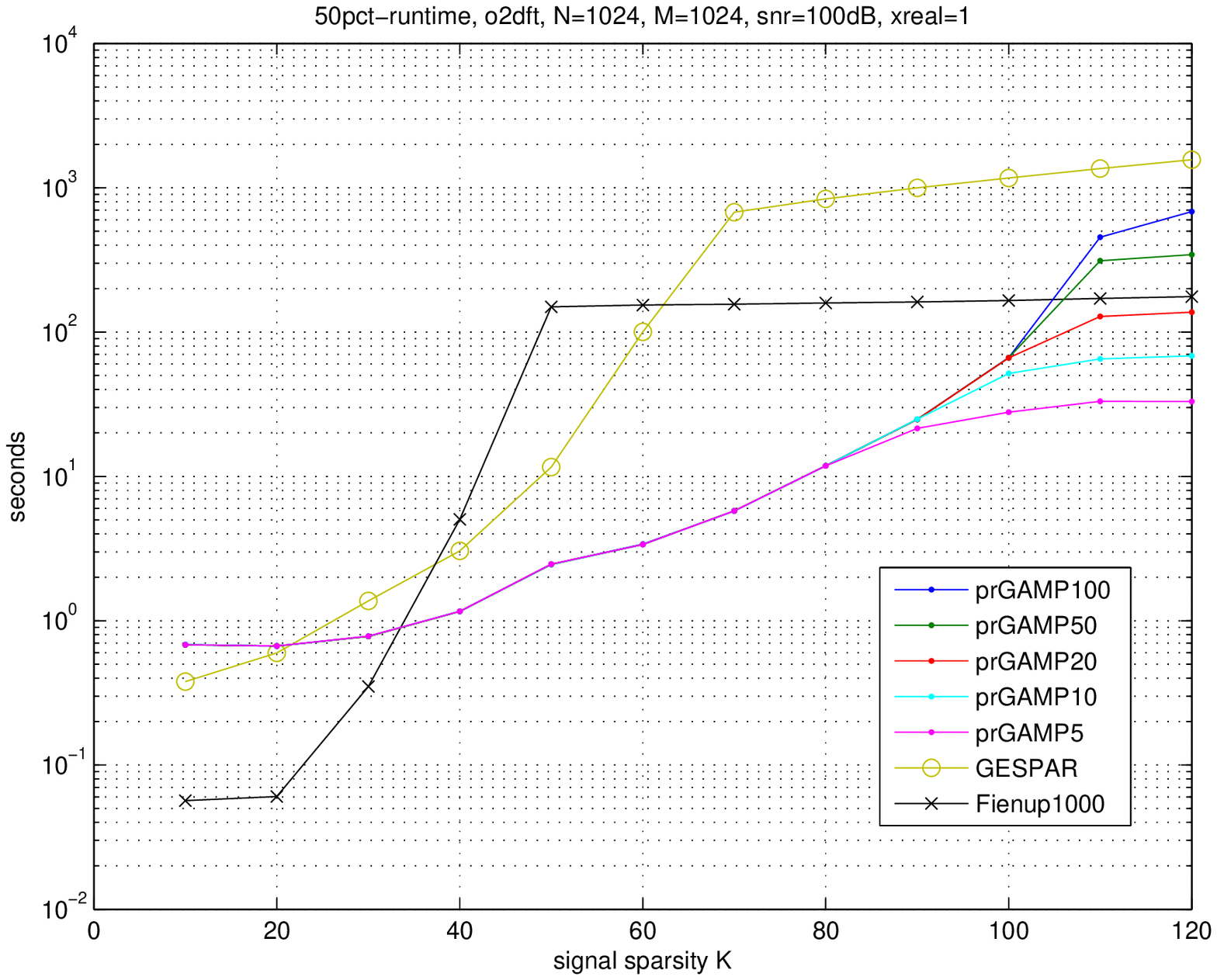}
	{Median runtime versus sparsity $K$ in the recovery of an $N\!=\!1024$-length real-valued signal from $M\!=\!1024$ 2D-Fourier intensities at $\SNR=100$dB. PR-GAMP-$A$ denotes PR-GAMP under a maximum of $A$ attempts.}
	{3.2}
	{\psfrag{signal sparsity K}[t][t][0.6]{\sf sparsity $K$} 
	 \psfrag{prGAMP200}[l][l][0.43]{\sf PR-GAMP-200} 
	 \psfrag{prGAMP100}[l][l][0.43]{\sf PR-GAMP-100} 
	 \psfrag{prGAMP50}[l][l][0.43]{\sf PR-GAMP-50} 
	 \psfrag{prGAMP20}[l][l][0.43]{\sf PR-GAMP-20} 
	 \psfrag{prGAMP10}[l][l][0.43]{\sf PR-GAMP-10} 
	 \psfrag{prGAMP5}[l][l][0.43]{\sf PR-GAMP-5} 
	 \psfrag{GESPAR}[l][l][0.43]{\sf GESPAR}
	 \psfrag{Fienup1000}[l][l][0.43]{\sf Fienup}
	 \psfrag{seconds}[b][b][0.6]{\sf seconds}
	 \psfrag{50pct-runtime, o2dft, N=1024, M=1024, snr=100dB, xreal=1}{}}

Next we consider 1D Fourier $\vec{A}$.
Again, we repeat an experiment from \cite{Shechtman:TSP:14}, where the measurement and signal lengths were fixed at $M=2N$ and the signal sparsity $K$ was varied.
For $N=1024$, \figref{success_1Dfft} shows the empirical success rate (over $200$ realizations) for PR-GAMP, GESPAR, and sparse Fienup, and \figref{time_2Dfft} shows the corresponding median runtimes.
From \figref{success_1Dfft}, we can see that PR-GAMP produced a significantly better phase-transition than GESPAR and sparse Fienup. 
Meanwhile, from \figref{time_1Dfft} we see that, for the challenging case of $K\geq 40$, PR-GAMP-20 had uniformly better runtime \emph{and} success rate than GESPAR and sparse Fienup.

\putFrag{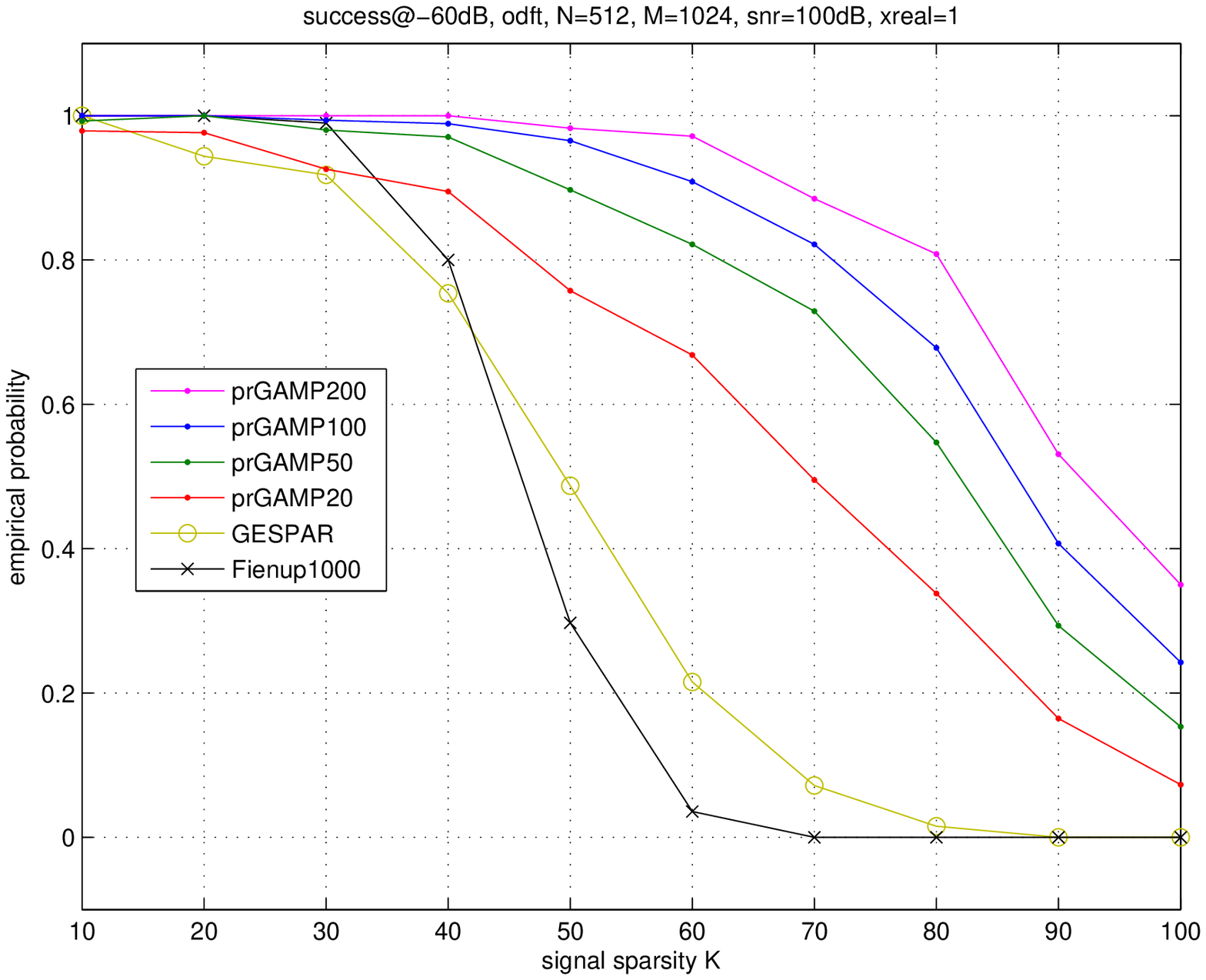}
	{Empirical success rate versus sparsity $K$ in the recovery of an $N\!=\!512$-length real-valued signal from $M\!=\!1024$ 1D-Fourier intensities at $\SNR=100$dB. PR-GAMP-$A$ denotes PR-GAMP under a maximum of $A$ attempts.}
	{3.2}
	{\psfrag{signal sparsity K}[t][t][0.6]{\sf sparsity $K$} 
	 \psfrag{prGAMP200}[l][l][0.43]{\sf PR-GAMP-200} 
	 \psfrag{prGAMP100}[l][l][0.43]{\sf PR-GAMP-100} 
	 \psfrag{prGAMP50}[l][l][0.43]{\sf PR-GAMP-50} 
	 \psfrag{prGAMP20}[l][l][0.43]{\sf PR-GAMP-20} 
	 \psfrag{GESPAR}[l][l][0.43]{\sf GESPAR}
	 \psfrag{Fienup1000}[l][l][0.43]{\sf Fienup}
	 \psfrag{empirical probability}[b][b][0.6]{\sf empirical success rate}
	 \psfrag{success@-60dB, odft, N=512, M=1024, snr=100dB, xreal=1}{}}

\putFrag{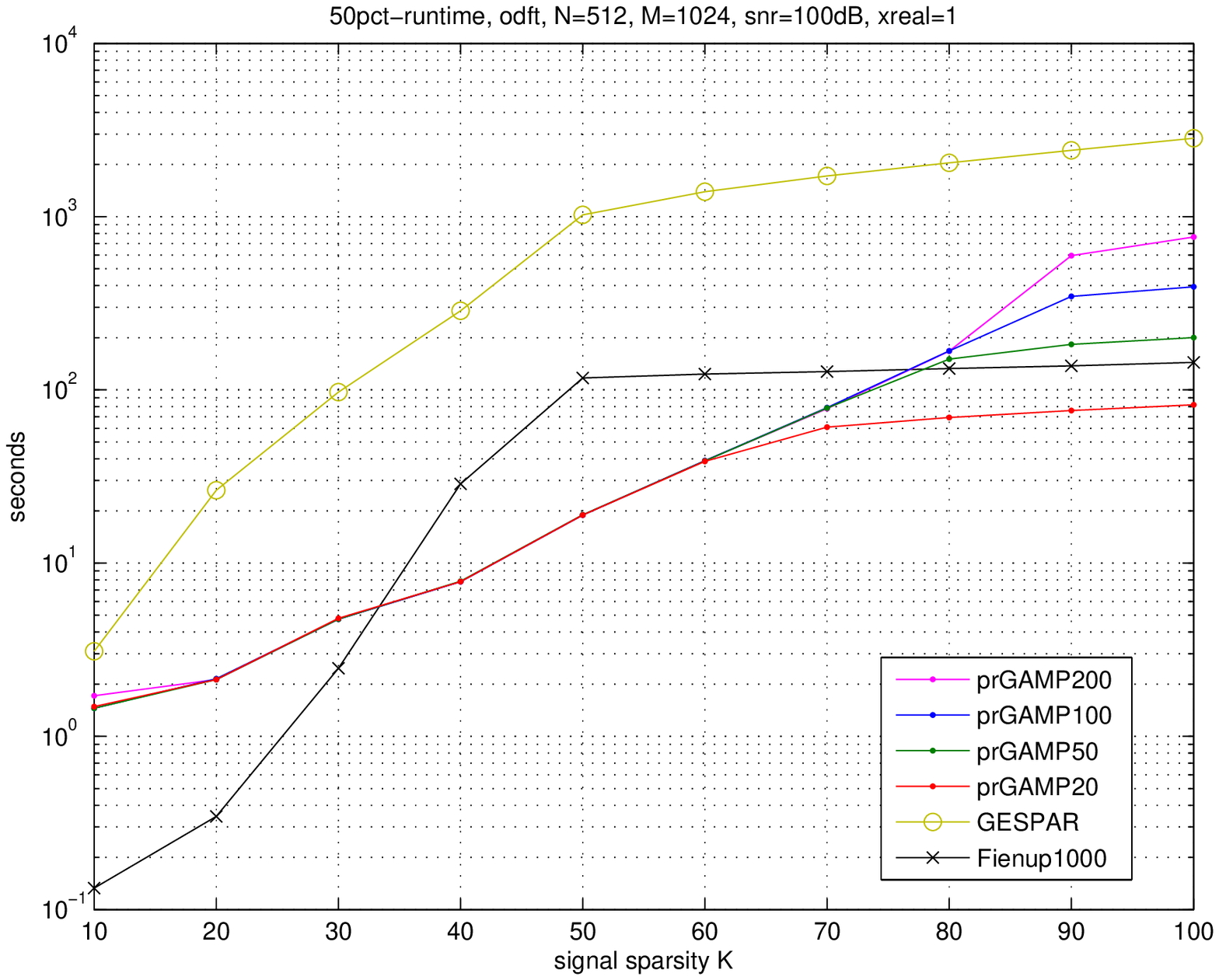}
	{Median runtime versus sparsity $K$ in the recovery of an $N\!=\!512$-length real-valued signal from $M\!=\!1024$ 1D-Fourier intensities at $\SNR=100$dB. PR-GAMP-$A$ denotes PR-GAMP under a maximum of $A$ attempts.}
	{3.2}
	{\psfrag{signal sparsity K}[t][t][0.6]{\sf sparsity $K$} 
	 \psfrag{prGAMP200}[l][l][0.43]{\sf PR-GAMP-200} 
	 \psfrag{prGAMP100}[l][l][0.43]{\sf PR-GAMP-100} 
	 \psfrag{prGAMP50}[l][l][0.43]{\sf PR-GAMP-50} 
	 \psfrag{prGAMP20}[l][l][0.43]{\sf PR-GAMP-20} 
	 \psfrag{prGAMP10}[l][l][0.43]{\sf PR-GAMP-10} 
	 \psfrag{prGAMP5}[l][l][0.43]{\sf PR-GAMP-5} 
	 \psfrag{GESPAR}[l][l][0.43]{\sf GESPAR}
	 \psfrag{Fienup1000}[l][l][0.43]{\sf Fienup}
	 \psfrag{seconds}[b][b][0.6]{\sf seconds}
	 \psfrag{50pct-runtime, odft, N=512, M=1024, snr=100dB, xreal=1}{}}

Comparing the results in this section to those in \secref{ptc}, we \textb{observe that the PR-GAMP, GESPAR, and Fienup algorithms had a much more difficult time with Fourier matrices $\vec{A}$ than with i.i.d matrices $\vec{A}$.
Similar observations were made \textb{in previous studies}, leading to proposals of randomized Fourier-based phase retrieval, e.g., using ``coded'' binary masks \cite{Candes:coded:13}.}
Also, we notice that the use of multiple restarts in PR-GAMP was much more important with Fourier $\vec{A}$ than it was with i.i.d $\vec{A}$.

\subsection{Practical image recovery with masked Fourier $\vec{A}$}	\label{sec:image}

\begin{figure}[h]
\centering
\psfrag{true}[B][B][0.7]{\sf original}
\psfrag{GAMP (-37.5dB)}[B][B][0.7]{\sf PR-GAMP (NMSE = \textb{-37.5} dB)}
\newcommand{\sz}{1.65in}
\includegraphics[width=\sz]{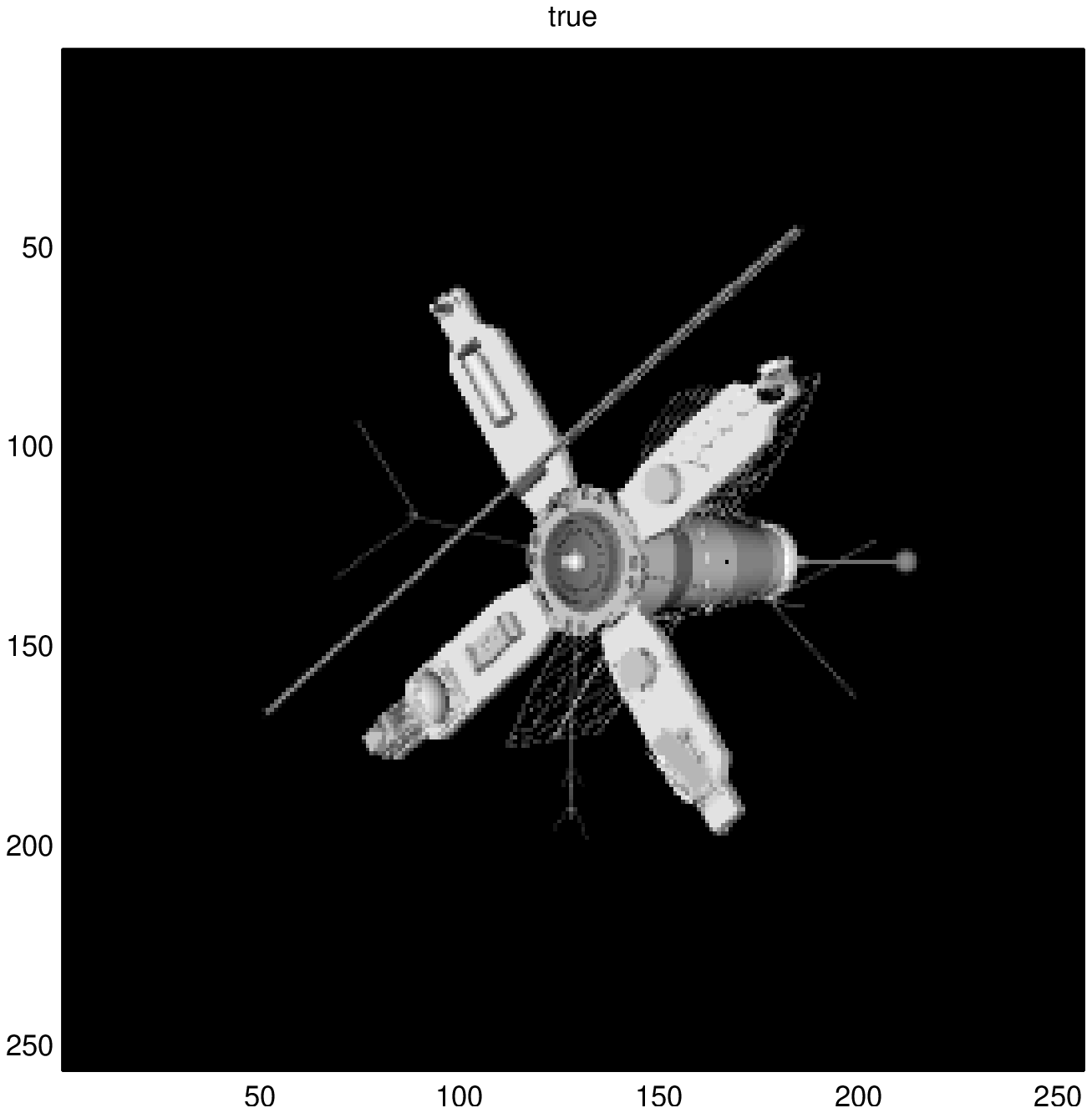}
\hfill
\includegraphics[width=\sz]{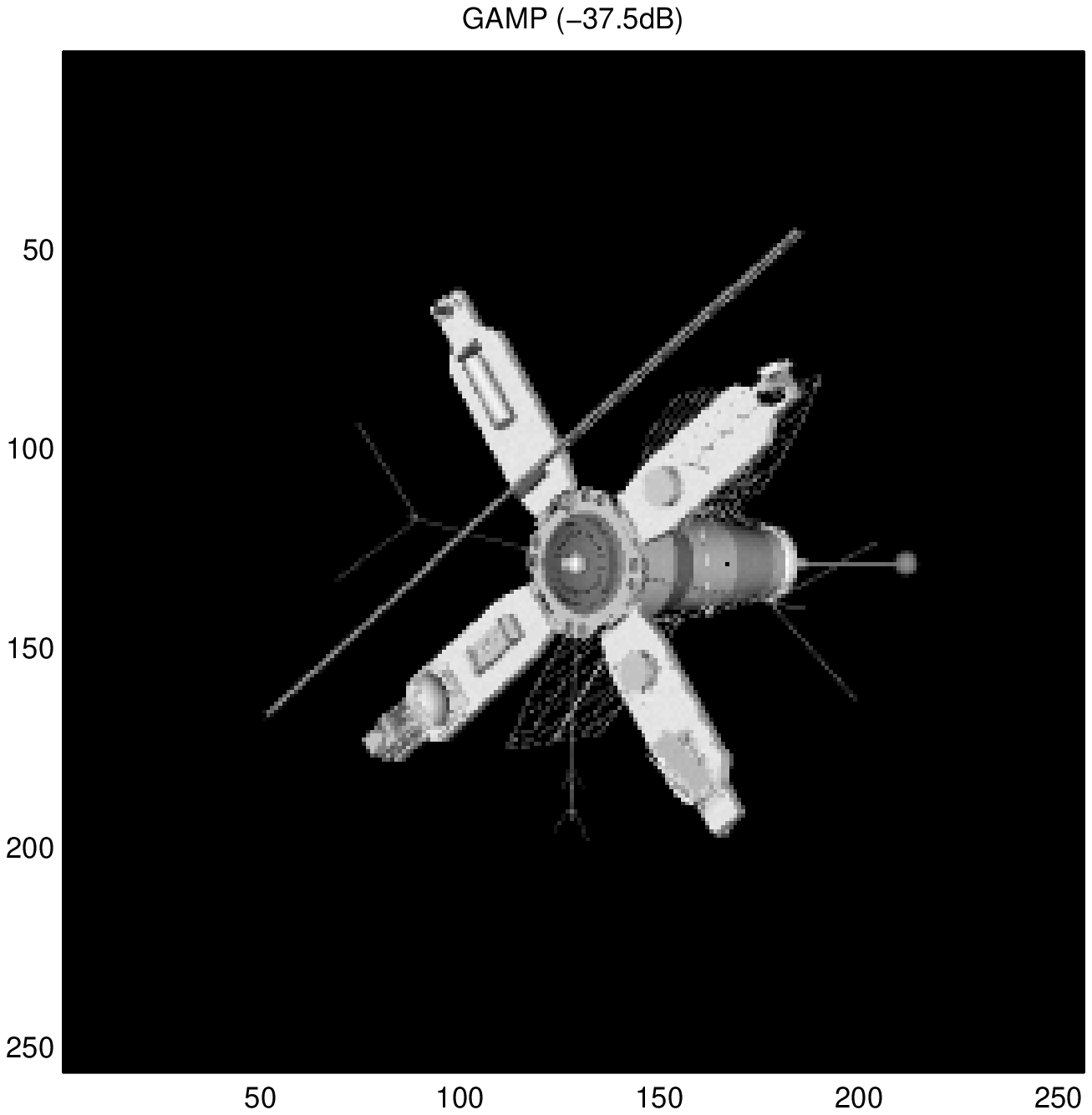}
\caption{Original image (left) and a typical PR-GAMP-recovery (right) from $M\!=\!N$ masked-Fourier intensity measurements at $\SNR\!=\!30$~dB, which took \textb{$1.8$} seconds.}
\label{fig:image}
\end{figure}

Finally, we demonstrate practical image recovery from compressed intensity measurements.  
For this experiment, the signal $\vec{x}$ was the $N=65536$-pixel grayscale image shown on the left of \figref{image}, which has a sparsity of $K=6678$. 
Since this image is real and non-negative, we ran PR-GAMP with a non-negative-real-BG prior \cite{Vila:CAMSAP:13}, as opposed to the BG prior \eqref{BG} used in previous experiments.

For the first set of experiments, we used a ``masked'' Fourier transformation $\vec{A}\in\Complex^{M\times N}$ of the form
\begin{equation}
  \vec{A} = \mat{\vec{J}_1\vec{FD}_1 \\ \vec{J}_2\vec{FD}_2 \\ \vec{J}_3\vec{FD}_3 \\ \vec{J}_4\vec{FD}_4} , \label{eq:masked}
\end{equation}
where 
$\vec{F}$ was a 2D DFT matrix of size $N\times N$,
$\vec{D}_i$ were diagonal ``masking'' matrices of size $N\times N$ with diagonal entries drawn uniformly at random from $\{0,1\}$, and
$\vec{J}_i$ were ``selection'' matrices of size $\frac{M}{4}\times N$ constructed from rows of the identity matrix drawn uniformly at random.
The matrices $\vec{D}_i$ and $\vec{J}_i$ help to ``randomize'' the DFT, and they circumvent unicity issues such as shift and flip ambiguities.
For phase retrieval, the use of image masks was discussed in \cite{Candes:coded:13}.
Note that, because $\vec{D}_i$ and $\vec{J}_i$ are sparse and $\vec{F}$ has a fast FFT-based implementation, the overall matrix $\vec{A}$ has a fast implementation.

To eliminate the need for the expensive matrix multiplications with the elementwise-squared versions of $\vec{A}$ and $\vec{A}\herm$, as specified in lines (S1) and (S6) of \tabref{step}, GAMP was run in ``uniform variance'' mode, meaning that $\{\nu^p_m(t)\}_{m=1}^M$ were approximated by $\nu^p(t)\defn \frac{1}{M}\sum_{m'=1}^M \nu_{m'}^p(t)$; similar was done with $\{\underline{\nu}^s_m(t)\}_{m=1}^M$, $\{\nu^r_n(t)\}_{n=1}^N$, and $\{\nu^x_n(t)\}_{n=1}^N$.
The result is that lines (S1)-(S2) in \tabref{step} become $\nu^p(t)=\beta \norm{\vec{A}}_F^2 \nu^x(t)/M + (1-\beta)\nu^p(t\!-\!1) = \alpha(t)$ and line (S6) becomes $\underline{\nu}^r(t) = \big(\norm{\vec{A}}_F^2 \underline{\nu}^s(t)/N\big)^{-1}$.

As before, the observations took the form $\vec{y}=|\vec{Ax} + \vec{w}|$, but now the noise variance was adjusted to yield a nontrivial $\SNR=30$~dB.
To demonstrate \emph{compressive} phase retrieval, only $M=N=65536$ intensity measurements were used.
Running PR-GAMP on $100$ problem realizations (each with different random $\vec{A}$ and $\vec{w}$, and allowing at most $10$ restarts per realization), \textb{we observed $\NMSE<-36$~dB for all $100$ realizations and a median runtime of only \textbf{\textb{5.9} seconds}}.
The right subplot in \figref{image} shows a typical PR-GAMP recovery.

For the second set of experiments, we ``blurred'' the masked-Fourier outputs to further randomize $\vec{A}$, which allowed us to achieve similar recovery performance using \emph{half} the intensity measurements, i.e., $M=\frac{N}{2}=32768$.
In particular, we used a linear transformation $\vec{A}\in\Complex^{M\times N}$ of the form
\begin{equation}
  \vec{A} = \mat{\vec{B}_1\vec{FD}_1 \\ \vec{B}_2\vec{FD}_2 }, 
  \label{eq:blurred}
\end{equation}
where $\vec{F}$ and $\vec{D}_i$ were as before\footnote{%
Here, since we used only two masks, we ensured invertibility by constructing the diagonal of $\vec{D}_1$ using exactly $N/2$ unit-valued entries positioned uniformly at random and constructing the diagonal of $\vec{D}_2$ as its complement, so that $\vec{D}_1+\vec{D}_2=\vec{I}$.}
and $\vec{B}_i$ were banded\footnote{%
Since each $\vec{B}_i$ was a wide matrix, its nonzero band was wrapped from bottom to top when necessary.} 
matrices of size $\frac{M}{2}\times N$ with $10$ nonzero i.i.d circular-Gaussian entries per column.
The use of blurring to enhance phase retrieval was discussed in \cite{Zheng:NP:13}.
As with \eqref{masked}, the $\vec{A}$ in \eqref{blurred} has a fast implementation. 
Running PR-GAMP as before on $100$ problem realizations at $\SNR=30$~dB, \textb{we observed $\NMSE<-28$~dB for all $100$ realizations and a median runtime of only \textbf{7.3 seconds}}.

To our knowledge, no existing algorithms are able to perform compressive phase retrieval on images of this size and sparsity with such high speed and accuracy.
To put our results in perspective, we recall the image recovery experiment in \cite{Shechtman:TSP:14}, which shows an example of GESPAR taking $80$ seconds to recover a $K=15$-sparse image whose support was effectively constrained to $N=225$ pixels from $M=38025$ 2D Fourier intensity measurements.
In contrast, \figref{image} shows PR-GAMP taking $\textb{1.8}$ seconds to recover a $K=6678$-sparse image whose support was constrained to $N=65536$ pixels from $M=65536$ masked 2D Fourier intensity measurements. 

\section{Conclusions}

In this paper, we proposed a novel approach to compressive phase retrieval based on the generalized approximate message passing (GAMP) algorithm.
Numerical results showed that the proposed PR-GAMP algorithm has excellent phase transition behavior, noise robustness, and runtime.
In particular, for successful recovery of synthetic $K$-sparse signals PR-GAMP requires approximately $4$ times the number of measurements as phase-oracle GAMP and achieves $\NMSE$ that is only $3$~dB worse than phase-oracle GAMP.
For recovery of a real-valued $65532$-pixel image from $32768$ pre-masked and post-blurred Fourier intensities, PR-GAMP \textb{returned $\NMSE<-28$~dB for all $100$ realizations and a median runtime of only 7.3 seconds}.
An extensive numerical comparison to the recently proposed CPRL, sparse-Fienup, and GESPAR algorithms \textb{suggests that PR-GAMP has} superior phase transitions and orders-of-magnitude faster runtimes at large $K$.


\appendices

\section{Output Thresholding Rules} \label{app:gout}
In this appendix, we derive the expressions \eqref{Ez|yp} and \eqref{Vz|yp} that are used to compute the functions $g\out$ and $g'\out$ defined in lines (D2) and (D3) of \tabref{gamp}.

To facilitate the derivations in this appendix,\footnote{The subscript ``$m$'' is omitted throughout this appendix for brevity.}
we first rewrite $p_{Y|Z}(y|z)$ in a form different from \eqref{pY|Z}.
In particular, recalling that---under our AWGN assumption---the noisy transform outputs $u=z+w$ are conditionally distributed as 
$p(u|z)=\mc{N}(u;z,\nu^w)$, 
we first transform $u=ye^{j \theta}$ from rectangular to polar coordinates to obtain
\begin{align}
  p(y,\theta|z)
  &= 1_{y\geq 0} 1_{\theta\in[0,2\pi)}\, \mc{N}(y e^{j\theta};z,\nu^w) \,y \label{eq:pytheta|z}
\end{align}
where $y$ is the Jacobian of the transformation, and then integrate out the unobserved phase $\theta$ to obtain
\begin{align}
  p_{Y|Z}(y|z)
  &= 1_{y\geq 0} \, y \int_0^{2\pi}\mc{N}(y e^{j\theta};z,\nu^w) 
  	\, d\theta , 	\label{eq:py|z}
\end{align}

We begin by deriving the scaling factor
\begin{align}
\lefteqn{
C(y,\nu^w,\hat{p},\nu^p) 
\defn \int_{\Complex} p_{Y|Z}(y|z)\, \mc{N}(z;\hat{p},\nu^p) dz 
}\nonumber\\
&= y \,1_{y\geq 0} 
	\int_0^{2\pi} 
	\int_{\Complex} 
	\mc{N}(y e^{j\theta};z,\nu^w) 
	\mc{N}(z;\hat{p},\nu^p) 
	dz 
	d\theta \\
&= y \,1_{y\geq 0} 
	\int_0^{2\pi}
	\mc{N}(y e^{j\theta};\hat{p},\nu^w+\nu^p) 
	d\theta ,	\label{eq:C1}
\end{align}
where we used the Gaussian-pdf multiplication rule\footnote{$\mc{N}(z;a,A)\mc{N}(z;b,B) \!=\! \mc{N}\Big(z; \frac{\frac{a}{A} + \frac{b}{B}}{\frac{1}{A} + \frac{1}{B}},\frac{1}{\frac{1}{A} + \frac{1}{B}}\Big)\mc{N}(a;b,A\!+\!B).$}
in \eqref{C1}.
Noting the similarity between \eqref{C1} and \eqref{py|z},
the equivalence between \eqref{py|z} and \eqref{pY|Z} implies that
\begin{align}
\lefteqn{ C(y,\nu^w,\hat{p},\nu^p) }\nonumber\\
&= \frac{2y}{\nu^w+\nu^p}
     \exp\Big(-\frac{y^2+|\hat{p}|^2}{\nu^w+\nu^p}\Big)
     I_0\Big(\frac{2y|\hat{p}|}{\nu^w+\nu^p}\Big)
     \,1_{y\geq 0} . \label{eq:C}
\end{align}
In the sequel, we make the practical assumption that $y>0$, allowing us to drop the indicator ``$1_{y\geq 0}$'' and invert $C$.

Next, we derive the conditional mean
\begin{align}
\lefteqn{ 
\E_{Z|Y,P}\{ Z|y,\hat{p};\nu^p\} 
} \label{eq:Ez|yp1} \\
&= C(y,\nu^w,\hat{p},\nu^p)^{-1}
\int_{\Complex} z\, p_{Y|Z}(y|z;\nu^w) \mc{N}(z;\hat{p},\nu^p) dz .
\nonumber
\end{align}
Plugging \eqref{py|z} into \eqref{Ez|yp1} and applying the Gaussian-pdf multiplication rule,
\begin{align}
\lefteqn{
\E_{Z|Y,P}\{ Z|y,\hat{p};\nu^p\} 
}\nonumber\\
&= C^{-1} y 
  \int_0^{2\pi} 
  \int_{\Complex} 
  z\, \mc{N}(z;ye^{j\theta},\nu^w) 
  \mc{N}(z;\hat{p},\nu^p) 
  dz d\theta \\
&= C^{-1} y 
  \int_0^{2\pi} 
  \int_{\Complex} 
  z\, \mc{N}(z;\tfrac{ye^{j\theta}/\nu^w + \hat{p}/\nu^p}{1/\nu^w+1/\nu^p},
  	\tfrac{1}{1/\nu^w+1/\nu^p}) 
\nonumber\\&\quad\times
  \mc{N}(ye^{j\theta};\hat{p},\nu^w\!+\!\nu^p) 
  dz d\theta \\
&= C^{-1} y
  \int_0^{2\pi} 
  \tfrac{ye^{j\theta}/\nu^w + \hat{p}/\nu^p}{1/\nu^w+1/\nu^p}
  \mc{N}(ye^{j\theta};\hat{p},\nu^w\!+\!\nu^p) 
  d\theta \\
&= \tfrac{y/\nu^w}{1/\nu^w+1/\nu^p}
  C^{-1} y\int_0^{2\pi} 
  e^{j\theta}\mc{N}(ye^{j\theta};\hat{p},\nu^w\!+\!\nu^p) 
  d\theta 
\nonumber\\&\quad + 
  \tfrac{\hat{p}/\nu^p}{1/\nu^w+1/\nu^p} 
  C^{-1} y \int_0^{2\pi} 
  \mc{N}(ye^{j\theta};\hat{p},\nu^w\!+\!\nu^p) 
  d\theta \\
&= \tfrac{y}{\nu^w/\nu^p+1}
  C^{-1} y \int_0^{2\pi} 
  e^{j\theta}\mc{N}(ye^{j\theta};\hat{p},\nu^w\!+\!\nu^p) 
  d\theta 
\nonumber\\&\quad+
  \tfrac{\hat{p}}{\nu^p/\nu^w+1} . 	\label{eq:Ez|yp2}
\end{align}
Expanding the $\mc{N}$ term, the integral in \eqref{Ez|yp2} becomes  
\begin{align}
\lefteqn{
  \int_0^{2\pi} 
  e^{j\theta}\mc{N}(ye^{j\theta};\hat{p},\nu^w\!+\!\nu^p) 
  d\theta 
}\nonumber\\
  &= \tfrac{1}{\pi(\nu^w+\nu^p)}
    \exp\big(-\tfrac{y^2+|\hat{p}|^2}{\nu^w+\nu^p}\big)
\nonumber\\&\quad\times
    \int_0^{2\pi} 
    e^{j\theta} 
    \exp\big(\tfrac{2y |\hat{p}|}{\nu^w+\nu^p}\cos(\theta-\psi)\big)
    d\theta \\
  &= \tfrac{1}{\pi(\nu^w+\nu^p)}
    \exp\big(-\tfrac{y^2+|\hat{p}|^2}{\nu^w+\nu^p}\big)
\nonumber\\&\quad\times
    e^{j\psi}\int_0^{2\pi} 
    e^{j\theta'} 
    \exp\big(\tfrac{2y |\hat{p}|}{\nu^w+\nu^p}\cos(\theta')\big)
    d\theta' 
    	\label{eq:Ez|yp3} \\
  &= \frac{2e^{j\psi}}{\nu^w+\nu^p}
    \exp\Big(-\frac{y^2+|\hat{p}|^2}{\nu^w+\nu^p}\Big)
    I_1\Big(\frac{2y |\hat{p}|}{\nu^w+\nu^p}\Big)
    	\label{eq:Ez|yp4}
\end{align}
where $\psi$ denotes the phase of $\hat{p}$, and
where the integral in \eqref{Ez|yp3} was resolved using the expression in 
\cite[9.6.19]{Abramowitz:Book:64}.
Plugging \eqref{Ez|yp4} into \eqref{Ez|yp2} gives
\begin{align}
\lefteqn{
\E_{Z|Y,P}\{ Z|y,\hat{p};\nu^p\} 
}\nonumber\\
  &= \frac{\hat{p}}{\nu^p/\nu^w+1} + \frac{y e^{j\psi}}{\nu^w/\nu^p+1}
    \frac{I_1\big(\frac{2y |\hat{p}|}{\nu^w+\nu^p}\big)}
         {I_0\big(\frac{2y |\hat{p}|}{\nu^w+\nu^p}\big)} ,
    	\label{eq:Ez|yp5}
\end{align}
which agrees with \eqref{Ez|yp}.

Finally, we derive the conditional covariance
\begin{align}
\lefteqn{ 
\var_{Z|Y,P}\{ Z|y,\hat{p};\nu^p\} 
}\nonumber\\
&= 
C(y,\nu^w,\hat{p},\nu^p)^{-1}
\int_{\Complex} |z|^2\, p_{Y|Z}(y|z;\nu^w) \mc{N}(z;\hat{p},\nu^p) dz
\nonumber\\&\quad 
- |\E_{Z|Y,P}\{ Z|y,\hat{p};\nu^p\}|^2 .
	\label{eq:Vz|yp1}
\end{align}
Focusing on the first term in \eqref{Vz|yp1}, if we plug in \eqref{py|z} and apply the Gaussian-pdf multiplication rule, we get
\begin{align}
\lefteqn{ 
C(y,\nu^w,\hat{p},\nu^p)^{-1}
\int_{\Complex} |z|^2\, p_{Y|Z}(y|z;\nu^w) \mc{N}(z;\hat{p},\nu^p) dz
}\nonumber\\
&= C^{-1} y 
  \int_0^{2\pi} 
  \int_{\Complex} 
  |z|^2\, 
  \mc{N}(z;\tfrac{ye^{j\theta}/\nu^w + \hat{p}/\nu^p}{1/\nu^w+1/\nu^p},
  	\tfrac{1}{1/\nu^w+1/\nu^p}) dz
\nonumber\\&\quad\times
  \mc{N}(ye^{j\theta};\hat{p},\nu^w\!+\!\nu^p) 
  d\theta \\
&= C^{-1} y
  \int_0^{2\pi} 
  \Big(\big|\tfrac{ye^{j\theta}/\nu^w + \hat{p}/\nu^p}{1/\nu^w+1/\nu^p}\big|^2
  + \tfrac{1}{1/\nu^w+1/\nu^p}\Big)
\nonumber\\&\quad\times
  \mc{N}(ye^{j\theta};\hat{p},\nu^w\!+\!\nu^p) 
  d\theta \\
&= C^{-1} y
  \int_0^{2\pi} 
  \tfrac{|y|^2/(\nu^w)^2 + |\hat{p}|^2/(\nu^p)^2 
  + 2y|\hat{p}|/(\nu^w\nu^p)\real\{e^{j(\theta-\psi)}\}}
  {(1/\nu^w+1/\nu^p)^2}
\nonumber\\&\quad\times
  \mc{N}(ye^{j\theta};\hat{p},\nu^w\!+\!\nu^p) 
  d\theta 
  + \tfrac{1}{1/\nu^w+1/\nu^p} \\
&= \tfrac{|y|^2/(\nu^w)^2 + |\hat{p}|^2/(\nu^p)^2}{(1/\nu^w+1/\nu^p)^2}
  + \tfrac{1}{1/\nu^w+1/\nu^p} 
  + \tfrac{2y|\hat{p}|/(\nu^w\nu^p)}{(1/\nu^w+1/\nu^p)^2}
\nonumber\\&\quad \times
  C^{-1} y
  \real \Big\{ e^{-j\psi} \int_0^{2\pi} 
  e^{j\theta}
  \mc{N}(ye^{j\theta};\hat{p},\nu^w\!+\!\nu^p) 
  d\theta \Big\} \\
&= \tfrac{|y|^2/(\nu^w)^2 + |\hat{p}|^2/(\nu^p)^2}{(1/\nu^w+1/\nu^p)^2}
  + \tfrac{1}{1/\nu^w+1/\nu^p} 
  + \tfrac{2y|\hat{p}|/(\nu^w\nu^p)}{(1/\nu^w+1/\nu^p)^2}
\nonumber\\&\quad \times
  C^{-1} y
  \tfrac{2}{\nu^w+\nu^p}
    \exp\Big(-\tfrac{y^2+|\hat{p}|^2}{\nu^w+\nu^p}\Big)
    I_1\Big(\tfrac{2y |\hat{p}|}{\nu^w+\nu^p}\Big)
    	\label{eq:Vz|yp2} \\
&= \tfrac{|y|^2/(\nu^w)^2 + |\hat{p}|^2/(\nu^p)^2}{(1/\nu^w+1/\nu^p)^2}
  + \tfrac{1}{1/\nu^w+1/\nu^p} 
  + \tfrac{2y|\hat{p}|/(\nu^w\nu^p)}{(1/\nu^w+1/\nu^p)^2}
\nonumber\\&\quad \times
    I_1\Big(\tfrac{2y |\hat{p}|}{\nu^w+\nu^p}\Big)/
    I_0\Big(\tfrac{2y |\hat{p}|}{\nu^w+\nu^p}\Big)
    	\label{eq:Vz|yp3}
\end{align}
where \eqref{Vz|yp2} used \eqref{Ez|yp4} and \eqref{Vz|yp3} used \eqref{C}.
By plugging \eqref{Vz|yp3} back into \eqref{Vz|yp1}, we obtain the expression given in \eqref{Vz|yp}.

\section{Post-intensity Noise Models} \label{app:gout2}

\textb{
In this appendix, we consider the $g\out$ and $g'\out$ functions (defined in lines (D2) and (D3) of \tabref{gamp}) for the post-intensity noise model \eqref{ym2} under generic $q(\cdot)$ and $p_W$.
}

\textb{
Following the procedure in \appref{gout}, we begin by examining the scaling factor
\begin{align}
\lefteqn{
C(y,\hat{p},\nu^p) 
\defn \int_{\Complex} p_{Y|Z}(y|z)\, \mc{N}(z;\hat{p},\nu^p) dz 
}\nonumber\\
&= \int_0^\infty \!\!\! 
        p_W(y-q(r))
        \bigg(
	\int_0^{2\pi} 
	\mc{N}(r e^{j\phi};\hat{p},\nu^p) 
	d\phi
	\bigg) r\, dr \label{eq:C1_} \\
&= \frac{2}{\nu^p} \!
        \int_0^\infty \!\!\!\! 
        r\, p_W(y-q(r))
        \exp\Big(\!\!-\!\tfrac{r^2 + |\hat{p}|^2}{\nu^p}\Big)
        I_0\Big(\tfrac{2 r |\hat{p}|^2}{\nu^p}\Big)
	dr \label{eq:C_},
\end{align}
where for \eqref{C1_} we used the rectangular-to-polar transformation $z=r e^{j\phi}$ with $r\geq 0$ and $\phi\in[0,2\pi)$, and for \eqref{C_} we used the Rician result \eqref{pY|Z}.
}

\textb{
Next we examine the conditional mean defined in \eqref{Ez|yp1}.
Plugging \eqref{pY|Z2} into \eqref{Ez|yp1} and transforming from rectangular to polar coordinates, we get
\begin{align}
\lefteqn{
\E_{Z|Y,P}\{ Z|y,\hat{p};\nu^p\} 
}\nonumber\\
&= \frac{1}{C} 
  \int_0^\infty \!\!\! 
  r\, p_W(y-q(r)) 
  \bigg(
  \int_0^{2\pi} e^{j\phi} \mc{N}(re^{j\phi};\hat{p},\nu^p) d\phi 
  \bigg) r\, dr \\
&= \frac{2e^{j\psi}}{C \nu^p} \!\!
        \int_0^\infty \!\!\!\!
        r^2 p_W(y-q(r))
        \exp\Big(\!\!-\!\tfrac{r^2 + |\hat{p}|^2}{\nu^p}\Big)
        I_1\Big(\tfrac{2 r |\hat{p}|^2}{\nu^p}\Big)
	dr ,
  \label{eq:Ez|yp_}
\end{align}
where \eqref{Ez|yp_} used the result from \eqref{Ez|yp4}.
}

\textb{
Finally we examine the conditional covariance \eqref{Vz|yp1},
and in particular the first term in \eqref{Vz|yp1}, which now becomes
\begin{align}
\lefteqn{ 
\frac{1}{C}
\int_{\Complex} |z|^2\, p_{Y|Z}(y|z) \mc{N}(z;\hat{p},\nu^p) dz
}\nonumber\\
&= \frac{1}{C} 
  \int_0^\infty \!\!\!
  r^2
  p_W(y-q(r))
  \bigg(
  \int_0^{2\pi} \mc{N}(re^{j\phi};\hat{p},\nu^p) d\phi
  \bigg)
  r\, dr\\
&= \frac{2}{C \nu^p} \!
        \int_0^\infty \!\!\!\! 
        r^3 p_W(y-q(r))
        \exp\Big(\!\!-\!\tfrac{r^2 + |\hat{p}|^2}{\nu^p}\Big)
        I_0\Big(\tfrac{2 r |\hat{p}|^2}{\nu^p}\Big)
	dr \label{eq:Vz|yp_},
\end{align}
where \eqref{Vz|yp_} used a computation similar to \eqref{C_}.
Further simplification of the above expressions requires specification of $q(\cdot)$ and $p_W$.
}

\section{EM Update for Noise Variance} \label{app:EM}

\textb{%
In this appendix, we derive the EM update \eqref{EM} of the noise variance.
Our proposed is based on the use of GAMP's posterior approximation $b_{\vec{X}}(\vec{x})$ in place of the true posterior distribution $p_{\vec{X}|\vec{Y}}(\vec{x}|\vec{y})$ in \eqref{EM0}. 
At GAMP iteration $t$, 
$b_{\vec{X}}(\vec{x})=\prod_{n=1}^N b_{X_n\!}(x_n)$ for
\begin{align}
b_{X_n\!}(x)
&\defn p_{X_n\!}(x) \, 
\mc{N}\big(x;\hat{r}_n(t),\nu^r_n(t)\big) B_n^{-1}
\label{eq:bx} \\
B_n
&\defn \int_{\Complex} p_{X_n\!}(x) \, 
\mc{N}\big(x;\hat{r}_n(t),\nu^r_n(t)\big) dx ,
\end{align}
which also appears in line (D4) of \tabref{gamp}.
}

\textb{%
Under the posterior approximation $b_{\vec{X}}$ and large i.i.d $\vec{A}$, \cite{Krzakala:ISIT:14} claims that the negative log likelihood $-\ln p(\vec{y};\nu^w)$ is well approximated by a Bethe free entropy of the form\footnote{Note that the Bethe free entropy expressions in this paper are stated for the complex-valued case.} 
\begin{align}
\lefteqn{J\big(\nu^w;\hvec{r}(t),\vec{\nu}^r(t)\big)}\nonumber\\
&= 
D\big(b_{\vec{X}}\big\|p_{\vec{X}}\big) 
+ D\big(b_{\vec{Z}}\big\|p_{\vec{Y}|\vec{Z}}(\vec{y}|\cdot;\nu^w)\Gamma^{-1}\big) 
\nonumber\\&\quad
+ \sum_{m=1}^M\bigg(\frac{\var\{Z_m|b_{Z_m\!}\}}{\nu^p_m}+\ln(\pi\nu^p_m)\bigg)
\label{eq:Bethe} .
\end{align}
In \eqref{Bethe}, the first term measures the KL divergence of the prior $p_{\vec{X}}(\vec{x})\defn\prod_{n=1}^N p_{X_n}(x_n)$ from the approximated posterior $b_{\vec{X}}(\vec{x})$. 
%
The second term then measures the KL divergence of 
the pdf $p_{\vec{Y}|\vec{Z}}(\vec{y}|\vec{z};\nu^w)\Gamma^{-1}$ from $b_{\vec{Z}}(\vec{z})$, the GAMP-approximated posterior pdf on $\vec{z}$. 
Here, the scaling factor $\Gamma \defn \prod_{m=1}^M\Gamma_m$,
for
\begin{align}
\Gamma_m
\defn \int_{\Complex} p_{Y|Z}(y_m|z;\nu^w) dz ,
\end{align}
ensures that $p_{\vec{Y}|\vec{Z}}(\vec{y}|\vec{z};\nu^w)\Gamma^{-1}$ is a valid pdf over $\vec{z}\in\Complex^M$, and the approximate posterior takes the form $b_{\vec{Z}}(\vec{z})=\prod_{m=1}^M b_{Z_m\!}(z_m)$ for
\begin{align}
b_{Z_m\!}(z)
&\defn p_{Y|Z}(y_m|z;\nu^w) \, 
\mc{N}\big(z;\overline{p}_m,\nu^p_m\big) C_m^{-1}
\label{eq:bz} \\
C_m
&\defn \int_{\Complex} p_{Y|Z}(y_m|z;\nu^w) \, 
\mc{N}\big(z;\overline{p}_m,\nu^p_m\big) dz ,
\end{align}
which also appears in line (D1) of \tabref{gamp}.
Above, $(\ovec{p},\vec{\nu}^p)$ are ``fixed point'' values that are consistent with $\big(\hvec{r}(t),\vec{\nu}^r(t)\big)$ in the sense that 
\begin{align}
\nu^p_m 
&= \sum_{n=1}^N |a_{mn}|^2 
\underbrace{ \var\{X_n | b_{X_n}(\cdot;\hat{r}_n(t),\nu^r_n(t))\} }_{\displaystyle =\nu^x_n \text{~from~(R7)}}
\label{eq:pvarfix}\\
\hat{z}_m
&\defn \E\{Z_m | b_{Z_m\!}(\cdot;\overline{p}_m,\nu^p_m)\} \nonumber\\
&= \sum_{n=1}^N a_{mn} \underbrace{ \E\{X_n | b_{X_n}(\cdot;\hat{r}_n(t),\nu^r_n(t))\}  }_{\displaystyle =\hat{x}_n \text{~from~(R8)}}.
\label{eq:phatfix}
\end{align}
Whereas $\nu^p_m$ in \eqref{pvarfix} can be computed directly from $(\hvec{r}(t),\vec{\nu}^r(t))$, finding the $\overline{p}_m$ that solves \eqref{phatfix} may require numerical search, e.g., via Newton's method \cite{Vila:ICASSP:15}.
}

\textb{
Plugging \eqref{bz} into the second term of \eqref{Bethe} reveals
\begin{align}
\lefteqn{ D\big(b_{\vec{Z}}\big\|p_{\vec{Y}|\vec{Z}}(\vec{y}|\cdot;\nu^w)\Gamma^{-1}\big) }\nonumber\\
&= \int_{\Complex^M} b_{\vec{Z}}(\vec{z}) \ln \frac{b_{\vec{Z}}(\vec{z})}{p_{\vec{Y}|\vec{Z}}(\vec{y}|\vec{z}) \Gamma^{-1}} d\vec{z} \\
&= \sum_{m=1}^M \int_{\Complex} b_{Z_m\!}(z_m) \ln \frac{b_{Z_m\!}(z_m)}{p_{Y|Z}(y_m|z_m) \Gamma_m^{-1}} dz_m \\
&= \sum_{m=1}^M \int_{\Complex} b_{Z_m\!}(z) \ln \frac{\mc{N}\big(z;\overline{p}_m(t),\nu^p_m(t)\big) C_m^{-1}}{\Gamma_m^{-1}} dz \\
&= \sum_{m=1}^M \bigg( \ln \frac{\Gamma_m}{C_m} 
- \ln(\pi \nu^p_m) 
- \frac{\nu^z_m + |\hat{z}_m-\overline{p}_m|^2}{\nu^p_m}
\bigg) ,
\label{eq:Bethe2}
\end{align}
using the shorthand notation $\nu^z_m\defn \var\{Z_m|b_{Z_m\!}\}$.
Then plugging \eqref{Bethe2} into \eqref{Bethe} and canceling terms reveals
\begin{align}
\lefteqn{J\big(\nu^w;\hvec{r}(t),\vec{\nu}^r(t)\big)}
\label{eq:Bethe1}\\
&= D\big(b_{\vec{X}}\big\|p_{\vec{X}}\big)
-\sum_{m=1}^M \bigg( 
\ln \frac{C_m(\nu^w)}{\Gamma_m(\nu^w)}
+\frac{|\hat{z}_m-\overline{p}_m(\nu^w)|^2}{\nu^p_m}
\bigg) 
\nonumber ,
\end{align}
where the $\nu^w$ dependence of $\Gamma_m$, $C_m$, and $\overline{p}_m$ is made explicit.
Note that $\hat{z}_m$ and $\nu^p_m$ are completely determined by $(\hvec{r}(t),\vec{\nu}^r(t))$ via \eqref{pvarfix}-\eqref{phatfix}, and thus invariant to $\nu^w$, and
$D\big(b_{\vec{X}}\big\|p_{\vec{X}}\big)$ is by definition invariant to $\nu^w$.
When $p_{Y|Z}$ is Gaussian, the value of $\overline{p}_m(\nu^w)$ can be computed in closed form after which the resulting expression \eqref{Bethe1} simplifies \cite{Krzakala:JSM:12}.}

\textb{%
For non-Gaussian $p_{Y|Z}$, we propose the following EM update procedure.
For simplicity, we will assume that one EM update is performed per GAMP iteration, allowing us to write the EM iterations ``$[i]$'' as GAMP iterations ``$(t)$''.
Recalling \eqref{EM0}, we first run GAMP with $\hat{\nu^w}(t)$ to produce $(\hvec{r}(t),\vec{\nu}^r(t))$, the approximate posterior $b_{\vec{X}}(\vec{x})$ in \eqref{bx},
the corresponding $\big(\ovec{p}(\hat{\nu^w}(t)),\vec{\nu}^p\big)$ from \eqref{pvarfix}-\eqref{phatfix}, and finally the approximation of 
$-\E\big\{\ln p(\vec{y},\vec{x};\nu^w) \big| \vec{y}; \hat{\nu^w}(t) \big\}$ 
in \eqref{Bethe1}. 
However, to facilitate the minimization over $\nu^w$, we use 
\begin{align}
\lefteqn{\tilde{J}\big(\nu^w;\hvec{r}(t),\vec{\nu}^r(t),\hat{\nu^w}(t)\big)}
\label{eq:Bethe1b}\\
&\defn D\big(b_{\vec{X}}\big\|p_{\vec{X}}\big)
-\sum_{m=1}^M \bigg( 
\ln \frac{C_m(\nu^w)}{\Gamma_m(\nu^w)}
+\frac{|\hat{z}_m-\overline{p}_m(\hat{\nu^w}(t))|^2}{\nu^p_m}
\bigg) 
\nonumber 
\end{align}
in place of 
\eqref{Bethe1}, noting that the substitution of $\overline{p}_m(\nu^w)$ by $\overline{p}_m(\hat{\nu^w}(t))$ preserves the fixed point(s) of the EM procedure.
Finally, we assign the value of $\nu^w$ that minimizes \eqref{Bethe1b} to $\hat{\nu^w}(t\!+\!1)$.
The overall procedure is summarized by \eqref{EM}.%
}

\textb{%
For the $p_{Y|Z}$ in \eqref{pY|Z} used for PR-GAMP, 
it can be shown that $\Gamma_m(\nu^w)$ is invariant to $\nu^w$. 
Thus, $\hat{\nu^w}(t\!+\!1)=\arg\max_{\nu^w} \sum_{m=1}^M \ln C_m(\nu^w)$ for 
the $C_m(\nu^w)$ given in \eqref{C}.
We numerically compute the maximizing value.
}

\bibliographystyle{ieeetr}
\bibliography{macros_abbrev,books,misc,comm,sparse,machine,phase}
\end{document}